\documentclass[twocolumn,showpacs,showkeys,preprintnumbers,amsmath,amssymb,floatfix,superscriptaddress]{revtex4}
\usepackage{graphicx}
\usepackage{dcolumn}% Align table columns on decimal point
\usepackage{bm}% bold math
\newcommand{\bra}{\left\langle}
\newcommand{\ket}{\right\rangle}
\newcommand{\eref}[1]{Eq. (\ref{#1})}

\newcommand{\esref}[1]{Eqs. (\ref{#1})}

\newcommand{\fref}{Fig. \ref}
\newcommand{\Fref}{Figure \ref}
\newcommand{\fsref}{Figs. \ref}

\newcommand{\sref}{section \ref}

\newcommand{\cref}{chapter \ref}
\newcommand{\Cref}{Chapter \ref}
\newcommand{\appref}{appendix \ref}

\newcommand{\affA}{%
Centre for Atom Optics and Ultrafast Spectroscopy
and  ARC Centre of Excellence for Quantum-Atom Optics,
Swinburne University of Technology, Melbourne, Australia}
\newcommand{\affB}{%
ARC Centre of Excellence for Quantum Atom Optics, The University of Queensland, Brisbane, QLD 4072, Australia}
\begin{document}
\preprint{}
\title{Finite temperature correlations in the Bose-Hubbard model: application of the Gauge $P$ representation}
\author{Saeed Ghanbari}\affiliation{\affA}\email{sghanbari@swin.edu.au}
\author{Joel F. Corney}\affiliation{\affB}
\author{Tien D. Kieu}\affiliation{\affA}\altaffiliation[Permanent address:]{The Portland House Research Group Pty Ltd,
8 Collins Street, Melbourne, Australia 3000.}
\date{\today}
\begin{abstract}
We study ultracold Bose gases in periodic potentials as described by the Bose-Hubbard model. In 1D and at
finite temperature, we simulate ultracold Bose gases in imaginary time with the gauge
$P$ representation.  We study various quantities including the
Luttinger parameter $K$, which is important for locating the
boundaries of the Mott insulator lobes, and find a simple relation
for the kinetic energy part of the Bose-Hubbard Hamiltonian.
 We show
that for $J=0$, the stepwise pattern of
the average number of particles per lattice site versus the chemical potential
vanishes at temperatures above $T \approx 0.1 U$.  Also, at chemical potential $\mu=0.5 U$ and
temperature $T=0.5 U$ by increasing $J$, the relative value of the number fluctuation
decreases and approaches that of a coherent state.
\end{abstract}
\maketitle
We investigate quantum degenerate Bose gases at finite temperature
in the grand canonical ensemble~\cite{DrummondDeuar} and use the
Bose-Hubbard model which can describe the dynamics of ultracold
atoms in periodic potentials such as
optical~\cite{FisherM,Jaksch,Greiner} and magnetic
lattices~\cite{GhanbariKSH06,GhanbariKH07,Gerritsma07,Singh08,WhitlockGerritsma09,GhanbariSFMIMag09}.
A superfluid of ultracold bosons trapped in a periodic lattice
undergoes a superfluid to Mott insulator quantum phase transition if
the barrier height between the lattice sites is adiabatically
increased~\cite{FisherM,Jaksch,Greiner}.  In the Mott insulator
phase there is a fixed number of atoms per lattice site which may
have applications in quantum computation~\cite{FisherM,Jaksch}. The
Bose-Hubbard model is also important for the study of systems with
strongly correlated bosons~\cite{Bruder05Josephson}.

Ultracold bosons at zero temperature in the Bose-Hubbard model have been
studied using different methods such as Monte Carlo
simulations~\cite{BatrouniScalettar,Batrouni92,
KashurnikovRush96,capello07}, mean field
theory~\cite{SheshadriKrish,Oosten01,oktel}, Bethe-Ansatz
solution~\cite{Krauth91}, exact diagonalization~\cite{Elesin94},
strong-coupling expansions~\cite{Freericks96}, density-matrix
renormalization  group (DMRG) (infinite-size)~\cite{Kuhner98}, DMRG
(finite-size)~\cite{Kuhner00} and exact diagonalization plus
renormalization group~\cite{Kashurnikov96}.  DMRG~\cite{Kuhner00}
gives high precision results in only one dimensional many-body
problems~\cite{White92,Corboz07}.  Bose-Hubbard model at finite
temperature has been considered via mean-field
theory~\cite{SheshadriKrish,BuonsanteVezzani,lu:063615}, Monte Carlo
simulations of a quantum rotor model~\cite{cha:266406}, an $ab$
$initio$ stochastic method~\cite{PhysRevA.71.041601}, perturbative
DMRG~\cite{PhysRevA.70.013611} and
 slave particle techniques~\cite{Lu_Li_Yu_2006}.
So far, most of the finite temperature studies on the Bose-Hubbard
model have been based on perturbation theory or some approximations.

In this paper we use gauge $P$ representation which is an exact phase space
method based on a coherent state representation.
We promote the use of gauge
$P$ representation as an exact method to benchmark various
approximate methods and simplifying assumptions.  We evaluate the performance of this method for these imaginary time calculations of the Bose-Hubbard model to calculate correlations at finite temperature to connect between the different
limiting cases.
Applying a phase space method for a system with a
Hamiltonian written in the second quantized form, it is possible to
convert the master equation of the system (such as the Liouville-von
Neumann equation) into a differential
equation~\cite{GardinerZoller}. Phase space methods such as positive
$P$ representation~\cite{DrummondGardiner} and gauge $P$
representation~\cite{DeuarDrummond} can give accurate results which
their accuracy depends on the number of simulation trajectories.
Using the gauge $P$ representation, the second-order spatial
correlation function and also momentum distribution have been
calculated for an interacting 1D degenerate Bose gas in the
Bose-Hubbard model~\cite{DrummondDeuar}. We investigate ultracold atoms at finite
temperatures with the gauge $P$ representation~\cite{DeuarDrummond}
and open boundary conditions.  We choose  the gauge $P$
representation over other quantum phase space methods
 including the positive $P$
representation~\cite{DrummondGardiner} because in studying the
many-body physics problems with strongly correlated bosons, the
gauge $P$ representation gives more stable results  and also the
sampling error is reduced, compared with the positive $P$
representation~\cite{DeuarDrummondII,DeuarThesis}. Furthermore, the positive $P$/gauge $P$ is exact, whereas other phase-space methods are not.

We have verified the gauge $P$ technique by
comparisons with exact numerical and also analytical results in
simple cases~\cite{GhanbariThesis}.  Our simulation results are,
within the sampling error, in remarkable agreement with the
known limiting cases when either
the hopping matrix element $J$ or the onsite interaction $U$ is zero.  Therefore, they could be
 considered as a touchstone for testing the reliability of approximate
 techniques when both $J$ and $U$ are nonzero.  Nevertheless, with the
 present gauge in the gauge $P$ representation, because of increase in sampling error, simulation results for the the physical quantities in the Bose-Hubbard model are not precise at temperatures below $T=0.05U$, when both $J$ and $U$ have finite
  values.

In 1D, we simulate ultracold atoms in the Bose-Hubbard model with up to
 11 lattice sites and study the average number of
particles and coherence between lattice sites at finite
temperatures.  We show that for 11 lattice sites, the edge effects
are not very important and for a Bose-Hubbard model with 11 sites, we calculate
the Luttinger liquid parameter which is important for locating the
boundaries of the Mott insulator
lobes~\cite{GiamarchiMillis92,Giamarchi97,Glazman97}.

The Bose-Hubbard Hamiltonian is~\cite{FisherM,Jaksch}
\begin{equation}\label{BHHsecond}
{\hat H} = - J \sum_{<i,j>}^M \hat{a}_i^\dag \hat{a}_j  +
\frac{1}{2} U \sum_{i=1}^M \hat{n}_i (\hat{n}_i-1)+ \sum_{i=1}^M
\epsilon_i \hat{n}_i,
\end{equation}
where $M$  is the number of lattice sites and ${<i,j>}$ means that the
summation is taken over
 adjacent sites only.  Also, $\hat{a}_i^\dag$, ${\hat a}_i$ and $\hat{n}_i=\hat{a}_i^\dag
\hat{a}_i$ are
 creation, annihilation and number operators, respectively.  The canonical commutation relations for
$\hat{a}_i$ and $\hat{a}_j^\dag$ are $[\hat{a}_i,\hat{a}_j^\dag]=
\delta_{ij}$.  The hopping matrix element $J$ is defined by $ J= -
\int d^3{\bf x}w^* ({\bf x}-{\bf x}_i) [- {\hbar^2}/{2m}\nabla^2+
V_0({\bf x})]w({\bf x}-{\bf x}_j), $ where $V_0$ is a periodic
potential like the optical lattice (see~\cite{morsch06}, for a
review on optical lattices) or the magnetic
lattice~\cite{GhanbariKSH06,GhanbariKH07,GhanbariSFMIMag09}
potential. Wannier functions $w({\bf x}-{\bf x}_i)$ are localized
position eigenstates~\cite{Wannier}. The on-site interaction $ U =g
\int d^3 {\bf x}  |w({\bf x})|^4, $ where $g = {4 \pi a_s
\hbar^2}/{m}$ and $a_s$ and $m$ are the s-wave scattering
 length and mass of the bosonic atom, respectively.

At zero temperature, if $U{\gg}J$, the system is well into the Mott
insulator (MI) regime.
  For the Mott insulator with the commensurate filling
of  $n_i=n= N/M$, where $n_i$ are the number of particles  per
lattice site, the ground state of the system for $J=0$ is
described by $|\Psi_{{}_{MI}}\rangle_{{}_{J=0}}
=|n,\cdots,n\rangle$~\cite{LiuLu07}.
Atom-atom correlations
$C_i(r)$~\cite{Kuhner98,Kuhner00,Damski06} and density-density
correlations  $D_i(r)$~\cite{Damski06} are defined by $C_i(r)=\bra
{\hat a}^{\dagger}_i  {\hat a}^{}_{i+r}\ket$ and $D_i(r)=\bra{\hat
n}_i  {\hat n}_{i+r}\ket$, where $r$ is an integer and $0<r<M$. The
standard deviation of the number of particles~\footnote{Instead of
the long expression `the standard deviation of the number of
particles', for brevity, we may use the short term `standard
deviation'.} is $\Delta{n_i}= \sqrt{\langle {\hat n}^2_i \rangle -
\langle {\hat n}_i \rangle^2}$. For the ground state given by
$|\Psi_{{}_{MI}}\rangle_{{}_{J=0}} =|n,\cdots,n\rangle$, the
atom-atom correlations, the density-density correlations  and the
standard deviations at each site are $C_i(r)=0,
\hspace{0.2cm}D_i(r)=n^2$ and $\Delta{n_i} =0$~\cite{Damski06},
respectively. For a Mott insulator with $J\neq 0$, according to
first-order perturbation theory, we have $\Delta{n_i}
=J/U\sqrt{4dn(n+1)}$~\cite{plimak:013611}, where $d$ and $n$ are the
dimension of the system and the average number of particles in a
Mott insulator lobe, respectively.

At zero temperature, when the
on-site interaction $U$ is very small compared with  the hopping
matrix element $J$, the  system  is a superfluid (SF) and the ground
state for $U=0$ can be given as a coherent state $
|\Psi_{{}_{SF}}\rangle_{{}_{U=0}}
 %={1\over\sqrt{N!M^N}} \left( \sum_{i=1}^{M}\hat{a}_i^\dag \right)^N |0\rangle
 = \sqrt{N!/M^N}\sum_{\{ n_i\}}
  {{|n_1,\cdots,n_M\rangle}/\sqrt{n_1!\cdots n_M!}}$~\cite{Damski06},
where $\sum_i n_i=N$ is the total number of particles. Also, the sum
extends over all sets of occupation numbers $\{ n_i \}$ subject to
$0 \leq n_i \leq N$. For $M\gg 1$ and  commensurate filling $N/M=1$,
we obtain $C_i(r)=1, \hspace{0.2cm}D_i(r)=1$ and
$\hspace{0.2cm}\Delta{n_i} =1$~\cite{Damski06}. Therefore, for the
commensurate filling of $n=1$ we have the same density-density
correlations $D_i(r)=1$ for both the superfluid and  the Mott
insulator ground states.
% Henceforth, the parameters
%$J$, $\mu$, $\mu_e$ and $\mu_T$ are intended as $J/U$, $\mu/U$,
%$\mu_e/U$ and $\mu_T/U$, respectively.
\section{Formalism of the gauge $P$ representation and derivation of \^{I}to stochastic equations}
In this section, we perform some calculations in the positive $P$
representation and finally write the \^{I}to form of the Langevin equations in the gauge
$P$ representation. Then, in \sref{Simulations} we give the
simulation results. Also, in this section to obtain some general
expressions valid for both cases where $U\neq0$ and $U=0$, using the
Boltzmann constant $k_B=1$, we define dimensionless imaginary times
$\tau=U/T$ for $U\neq0$ and $\tau^\prime=J /T$ for $U=0$ and
$J\neq0$

Now, we consider the unnormalized density matrix
${{\hat\rho}}_u={\rm e}^{- ({\hat H}\tau-\mu {\hat N}\tau)}$, which
is in the grand canonical ensemble, and take its derivative with
respect to $\tau=U/T$~\cite{DrummondDeuar}
\begin{equation}
{{\partial{{\hat\rho}}_u}\over{\partial\tau}}= -{1\over
2}\left[{\hat H}- {\mu_e {\hat N}}, {\hat\rho}_u\right]_+
\end{equation}
where $[{\hat A},{\hat B}]_+ = {\hat A}{\hat B}+{\hat B}{\hat A}$ is
the anticommutator of ${\hat A}$ and ${\hat B}$.  Also, $\mu_e$ is defined by
\begin{equation}
\mu_e=\frac{\partial [\tau \mu(\tau)]}{\partial \tau}
\end{equation}
We now have
\begin{equation}\label{ro_derivative}
{{\partial{{\hat\rho}}_u}\over{\partial\tau}}= -{1\over
2}\left[{\hat H}^{\prime}({\hat{\bf a}},{\hat{\bf
a}^\dag}){\hat\rho}_u+ {\hat\rho}_u{\hat H}^{\prime}({\hat{\bf
a}},{\hat{\bf a}^\dag})\right]
\end{equation}
where
\begin{equation}
{\hat H}^{\prime}({\hat{\mathbf{{ a}}}},{\hat{\bf a}^\dag})={\hat
H}({\hat{\bf a}},{\hat{\bf a}^\dag})- \mu_e \sum_{i=1}^M {\hat{ \rm
a_i}^\dag}{\hat{\rm a_i}}
\end{equation}
According to \eref{rogauge}, in  the positive $P$ representation, we have
\begin{equation}\label{ro_u}
{{\hat\rho}}_u= \int P({\bm{\alpha}},{\bm{\beta}},\tau) {\hat{\Lambda} } d^{4M}
\vec{\lambda}
\end{equation}
Taking the derivative of  \eref{ro_u} with respect to $\tau$ and considering \eref{ro_derivative} and  \eref{ro_u}, we  obtain
\begin{eqnarray}
&~&\int{{{\partial} P({\bm{\alpha}},{\bm{\beta}},\tau)}\over{\partial\tau}}
 {\hat{\Lambda} } d^{4M} \vec{\lambda}=\nonumber \\
  &~&-{1\over 2}\int   P({\bm{\alpha}},{\bm{\beta}},\tau)  {\hat H}^{\prime}({\hat{\mathbf{{
a}}}},{\hat{\bf a}^\dag})  {\hat{\Lambda} } d^{4M} \vec{\lambda}\nonumber\\
 &~&+{1\over 2}  \int   P({\bm{\alpha}},{\bm{\beta}},\tau)  {\hat{\Lambda} } {\hat H}^{\prime}({\hat{\mathbf{{
a}}}},{\hat{\bf a}^\dag})   d^{4M} \vec{\lambda}
\end{eqnarray}

We  have, from \eref{identities}, ${\hat H_N}^{\prime}({\hat{\bf
a}},{\hat{\bf a}^\dag}) {\hat\Lambda}= {\hat
H_A}^{\prime}({\bm{\alpha}},{{{\bm{\beta}}}}+\partial_{\bm{\alpha}}){\hat\Lambda}$
and ${\hat\Lambda}{\hat H_N}^{\prime}({\hat{\bf a}},{\hat{\bf
a}^\dag})= {\hat
H_N}^{\prime}({\bm{\alpha}}+\partial_{\bm{\beta}},{{{\bm{\beta}}}}){\hat\Lambda}$
where
\begin{equation}\label{HNprime}
{\hat H_N}^{\prime}({\hat{\bf a}},{\hat{\bf a}^\dag}) = - J
\sum_{<i,j>}^M \hat{a}_i^\dag \hat{a}_j +  \frac{1}{2} U
\sum_{i=1}^M {{\hat a}^{\dag 2}}_i\hat{a}_i^2  -\mu_e \sum_{i=1}^M
{{\hat a}^\dag}_i \hat{a}_i
\end{equation}
\begin{equation}\label{HAprime}
{\hat H_A}^{\prime}({\hat{\bf a}},{\hat{\bf a}^\dag}) = - J
\sum_{<i,j>}^M \hat{a}_i \hat{a}_j^\dag +  \frac{1}{2} U
\sum_{i=1}^M \hat{a}_i^2{{\hat a}^{\dag 2}}_i  -\mu_e \sum_{i=1}^M
\hat{a}_i{{\hat a}^{\dag}}_i.
\end{equation}
So
\begin{equation}
\int {{\partial P({\bm{\alpha}},{\bm{\beta}},\tau)}\over{\partial\tau}}
 {\hat{\Lambda} } d^{4M} \vec{\lambda} =   \int   P({\bm{\alpha}},{\bm{\beta}},\tau)
{{{{\cal L}}^{(+)}_A}} {\hat{\Lambda} } d^{4M} \vec{\lambda}
 \end{equation}
where
\begin{equation}
{{{ {\cal L}}^{(+)}_A}} = -{1\over 2}\left[ {\hat
H_A}^{\prime}({\bm{\alpha}},{{{\bm{\beta}}}}+\partial_{\bm{\alpha}}) + {\hat
H_N}^{\prime}({\bm{\alpha}}+\partial_{\bm{\beta}},{{{\bm{\beta}}}}) \right]
 \end{equation}
According to \esref{HNprime} and (\ref{HAprime}), we also have
\begin{equation}\label{LABH}
{{{{\cal L}}^{(+)}_A}} = -{1\over 2}\left[ {\hat
H_N}^{\prime}({{{\bm{\beta}}}}+\partial_{\bm{\alpha}},{\bm{\alpha}}) + {\hat
H_N}^{\prime}({\bm{\alpha}}+\partial_{\bm{\beta}},{{{\bm{\beta}}}}) \right]
 \end{equation}
So
\begin{eqnarray}\label{AVD}
&~&\hspace{-0.5cm}{{{{\cal L}}^{(+)}_A}}=J\sum_{<i,j>}^M \alpha_i {\bf{\beta}}_j  -\frac{U}{2}
\sum_{i=1}^M \alpha_i^2 \beta_i^2  +\mu_e \sum_{i=1}^M
\alpha_i {\bf{\beta}}_i \nonumber \\
&+& \frac{J}{2} \sum_{<i,j>}^M (\alpha_i \partial_{\alpha_j}+\beta_i \partial_{\beta_j})
-\frac{U}{2}\sum_{i=1}^M n_i (\alpha_i \partial_{\alpha_i}+{\bf{\beta}}_i\partial_{\beta_i})
 \nonumber \\ &+& \frac{\mu_e}{2}   \sum_{i=1}^M
(\alpha_i \partial_{\alpha_i}   +{\bf{\beta}}_i\partial_{\beta_i})
 - \frac{U}{4} \sum_{i=1}^M  (\alpha_i^2{\partial^2_{\alpha_i}}+\beta_i^2{\partial^2_{\beta_i}} )
\end{eqnarray}
Comparing \eref{AVD} with \eref{AVD0}, we obtain
\begin{equation}\label{V}
V= J\sum_{i,j}^M \omega_{ij}\alpha_i {\bf{\beta}}_j  -\frac{U}{2}
\sum_{i=1}^M n_i^2  +\mu_e \sum_{i=1}^M
n_i
\end{equation}
\begin{equation}\label{A}
A^{(+)}_j=\frac{J}{2} \sum_{i=1}^{2M} {\omega_{ji}} \alpha^i - \frac{U}{2}n_j\alpha^j + \frac{\mu_e}{2}\alpha^j,  \quad j=1, 2, \cdots, 2M
\end{equation}
\begin{equation}\label{D}
D_{ij}=-\frac{U}{2}\delta_{ij} \alpha^{j2},  \quad i,j=1, 2, \cdots, 2M
\end{equation}
where $\omega_{ij}=\delta_{i,j-1}+\delta_{i-1,j}$,
$\omega_{i+M, j+M}=\omega_{ji}$ and $\omega_{i+M, j}=\omega_{i, j+M}=0$.
Also, the effective complex boson number is $n_i=n^\prime_i+ i n^{\prime\prime}_i=\alpha_i\beta_i=n_{i+M}$.
Considering \eref{D_BBT}, we have
\begin{equation}\label{B}
B_{ij}=i\sqrt{\frac{U}{2}}\delta_{ij} \alpha^{j}
\end{equation}
%\subsection{Fokker-Planck equation}
%\subsection{Langevin equations}
Choosing the gauge
$g_k=i\sqrt{\frac{U}{2}}(n^\prime_k-|n_k|)$~\cite{DeuarDrummond} and
considering \eref{B}, we can now calculate the \^{I}to equations and
convert them into the Stratonovich form of the Langevin  equations
(see \appref{ItoStra} for more calculations details). The
Stratonovich stochastic equations, which are a natural physical
choice~\cite{Gardiner} and more suitable for numerical solutions,
compared to the \^{I}to  stochastic equations, and also have
superior convergence
properties~\cite{DeuarDrummond,DrummondMortimer}, are
\begin{eqnarray}\label{StraOmega}
d\Omega^{(S)}&=& \Omega\left(V -\sum_{j=1}^{M} (g^2_j-i \frac{U}{2} n^{\prime\prime}_j)\right) d\tau \nonumber
\\&+& \Omega\sum_{k=1}^{2M} g_k dW_k
\end{eqnarray}
\begin{eqnarray}\label{StraAlphaj}
d\alpha^{j (S)}&=& \frac{J}{2} \sum_{i=1}^{2M} {\omega_{ji}} \alpha^id\tau - \frac{U}{2}(|n_j|+in^{\prime\prime}_j)\alpha^j d\tau \nonumber \\
&+& \frac{2\mu_e+U}{4}\alpha^jd\tau  + i \sqrt{\frac{U}{2}}\alpha^j dW_j
 \end{eqnarray}
 where the Wiener increments $dW_i$ have the property~\cite{DrummondGardiner,DeuarDrummondI}
\begin{equation}
\bra dW_i(\tau)dW_j(s)\ket_s =\delta_{ij}\delta(\tau-s)d\tau^2
\end{equation}
where $\bra \ket_s$, means  stochastic average.
\section{Simulations and comparisons in different limiting cases}\label{Simulations}
We simulate the 1D Bose-Hubbard model Stratonovich equations
\eref{StraAlphaj} using eXtensible Multi-Dimensional Simulator
(XMDS)~\footnote{Available at
http://www.physics.uq.edu.au/xmds/index. html}.  We use XMDS with the
semi-implicit interaction picture SIIP method.  We assume that at
$\tau=0$, there are $N$ particles on average, in a thermal
state~\cite{GardinerZoller} in the system. So the initial conditions
are
\begin{equation}\label{initial}
\alpha_j=\sqrt{N/2}(\xi_l+i\xi_m), \quad \beta_j=\alpha_j^\ast, \quad \Omega=1
\end{equation}
where $\xi_l$ and $\xi_m$ are random numbers with zero mean  and
standard deviation $1$ which can be realised by Gaussian random
numbers with zero mean  and standard deviation
$1$~\cite{GardinerZoller}.
\subsection{One-site model}\label{One-site model}
\begin{figure}% [t]
\begin{center}
$\begin{array}{c}
\includegraphics[angle=0,width=8.cm,height=5cm]{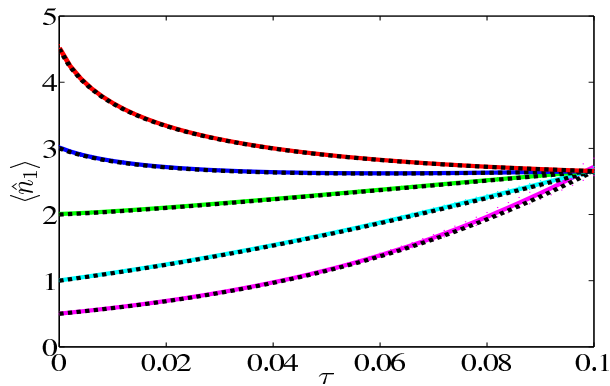}
\end{array}$
\caption[\small Simulation results: Independence of $\bra{\hat
n}_1\ket$ from the initial value of $n_0$.]{Solid lines show the
gauge $P$ numerical calculations of $\bra{\hat n}_1\ket$ versus
$\tau$ for $M=1$, $J=0$ and  $n_0=0.5, 1, 2, 3$ and $4.5$.  The
target chemical potential $\mu_{{}_T}$ is $0.5$. Solid, upper
dotted, and lower dotted lines show simulation results $\bra {\hat
n}_1\ket$, $\bra {\hat n}_1\ket + \sigma$ (sampling error) and $\bra
{\hat n}_1\ket - \sigma$, respectively. Black dashed lines show the
highly accurate numerical calculations using  a truncated
number-state basis.}\label{M1nos_Exact_and_Numerical}
\end{center}
\end{figure}

\Fref{M1nos_Exact_and_Numerical} compares the highly accurate
numerical calculations, using a truncated number state basis, with
the gauge $P$ representation simulations for
$M=1$~\cite{GhanbariThesis}. It shows the independence of the
expectation value of the number of particles $\bra{\hat n}_1\ket$ at
the target temperature $T=10\hspace{0.1cm}U$ (here,
$\tau_{{}_T}=0.1$) from the initial average number of particles
$n_0$, which is in good agreement with the numerical calculations
based on a truncated number-state basis. As the figure shows, for a
sufficiently large target imaginary time $\tau_{{}_T}$ (sufficiently
small temperature $T$) and a common value of the chemical potential
$\mu_{{}_T}$, $\bra{\hat n}_1\ket$  is independent of the initial
number of particles $n_0$~\cite{DeuarThesis}. According to the
figure, the expectation value of the number of particles  at
$\tau_{{}_T}=0.1$ is $2.66$, which is independent of the  different
initial values of $n_0=0.5, 1, 2, 3$ and $4.5$. Independence from
the $n_0$ is very useful for the phase space simulations, because
for different sets of the chemical potential and other parameters
such as $J$ and $U$ the sampling error and also the stability of the
simulations depends on $n_0$.

\begin{figure}%[t]
\hspace{-2cm}
\begin{center}
$\begin{array}{c}
\includegraphics[angle=0,width=8.cm,height=5cm]{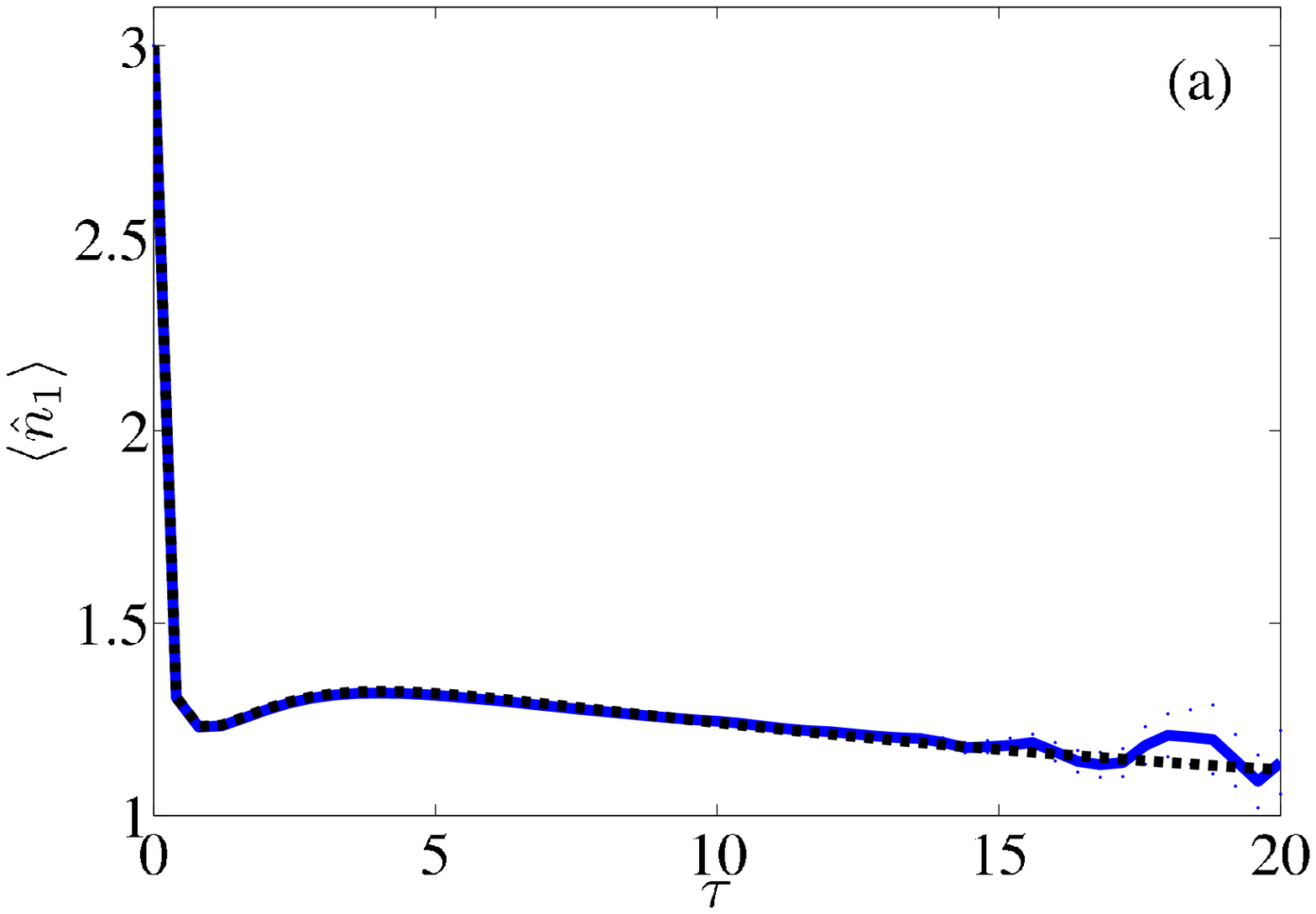}\\
\includegraphics[angle=0,width=8.cm,height=5cm]{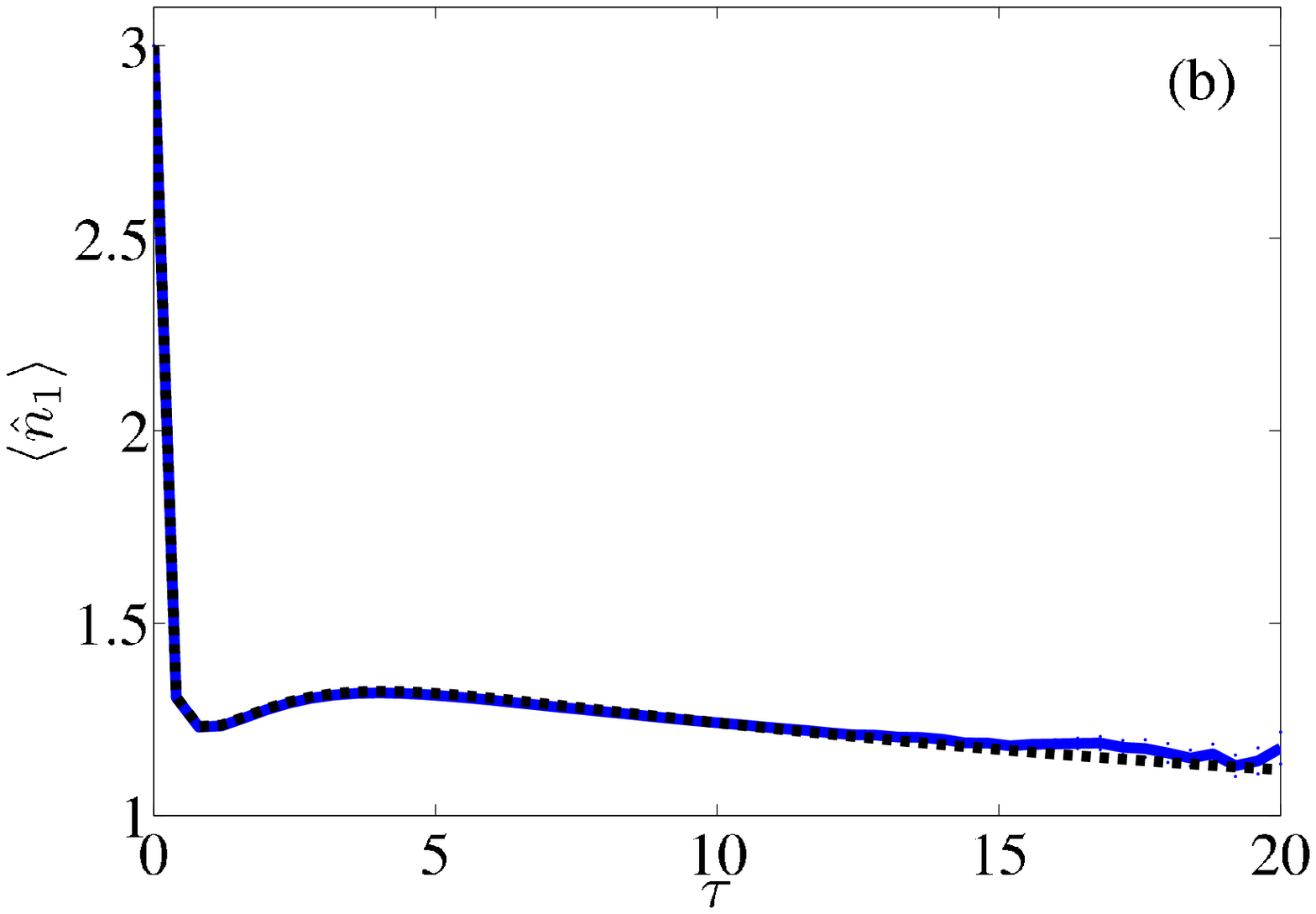}
\end{array}$
\caption[$\bra {\hat n}_1\ket$ versus $\tau$ for $M=1$, $n_0=3$,
$J=0$, $U=1.0$ and $\mu_{{}_T}=0.9$]{\small$\bra {\hat n}_1\ket$
versus $\tau$ for $M=1$, $n_0=3$, $J=0$, $U=1.0$, $d\tau=10^{-3}$,
$\mu_{{}_T}=0.9$ ($\mu_e =0.9144$) and  (a) $n_p=2 \times 10^9$
 and  (b) $n_p=2 \times 10^{10}$.  Conventions as in \fref{M1nos_Exact_and_Numerical}.}\label{M1_n1_tau_20_j0_muT_0_9}
\end{center}
\end{figure}

\begin{figure}%[t]%[tb]
\begin{center}
$\begin{array}{c}
\includegraphics[angle=0,width=8.cm,height=5cm]{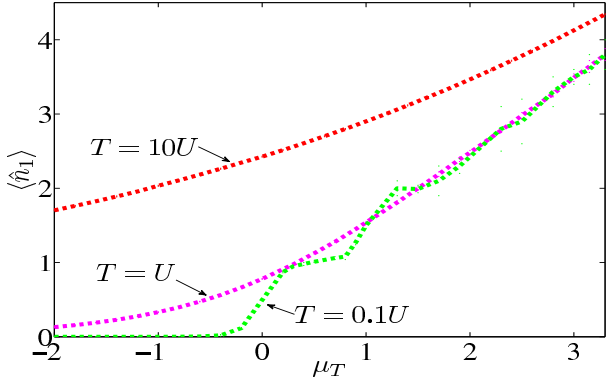}
\end{array}$
\caption[Simulation of $\bra{\hat n}_1\ket$ versus the chemical
potential $\mu_{{}_T}$]{\small  Gauge $P$ simulation of $\bra{\hat
n}_1\ket$ versus $\mu_{{}_T}$, for $M=1$, $n_0=1.2$, $J=0$, $U=1$
and three different values of $\tau_{{}_T}$.  Here again,
$\mu_{{}_T}$ is the chemical potential at the target temperature
$T$. As the temperatures is increased from $T=0.1\hspace{0.1cm}U$ to
$T=10\hspace{0.1cm}U$, the stepwise pattern of the average number of
particles versus the chemical potential, which is clearly seen below
$T_0=0.06\hspace{0.1cm}U$ as a truncated number-state basis shows,
vanishes.Dots around the diagram for $\tau_{{}_T}=10$ show the
sampling error.  For $T=U$ and $T=10 U$ sampling error is too small
to be seen.}\label{M1_n1_versus_muT_gauge_p}
\end{center}
\end{figure}

\Fref{M1_n1_tau_20_j0_muT_0_9} shows the average number of particles, $\bra {\hat n}_1\ket$, versus
the inverse temperature $\tau$ for the large value of
$\tau_{{}_T}=20$. There is  good agreement between the highly
accurate numerical values and the gauge $P$ simulation results.
\Fref{M1_n1_versus_muT_gauge_p} shows the simulation results for
$\bra{\hat n}_1\ket$ as a function of the chemical potential
$\mu_{{}_T}$.  By decreasing the temperature sampling error is
increased, specially at high values of the chemical potential.
It is possible to
reduce the sampling error by increasing the number of simulation
trajectories.  According to \fref{M1_n1_versus_muT_gauge_p}, the
stepwise pattern of the average number of particles as a function of
the chemical potential vanishes as the temperature is increased from
$T=0.1\hspace{0.1cm}U$ to $T=10\hspace{0.1cm}U$.  Because there is
no hopping matrix element $J$, this result is valid for an array of
$M$ lattice sites in 1D, 2D and 3D.  Therefore, for $J=0$, in a
system with $M$ lattice sites in 1D, 2D and 3D, the stepwise pattern
in the $\bra{\hat n}_i\ket$-$\mu_{{}_T}$ plot vanishes as the
temperature is increased from $T=0.1\hspace{0.1cm}U$ to
$T=10\hspace{0.1cm}U$.
For the Bose-Hubbard model, according to \esref{HNprime} and
(\ref{LABH}), we have
\begin{eqnarray}%\label{LABHe}
\hspace{-2cm}
&-&2{{{{\cal L}}^{(+)}_A}} = - J
\sum_{<i,j>}^M \alpha_i({{{\beta}}}_j+\partial_{\alpha_j}) \nonumber\\ &+&  \frac{1}{2} U
\sum_{i=1}^M \alpha_i^2 {({{{\beta}}}_i+\partial_{\alpha_i})}^2
-\mu_e \sum_{i=1}^M
\alpha_i ({{{\beta}}}_i+\partial_{\alpha_i})  \nonumber\\
&-&J\sum_{<i,j>}^M \beta_i ({\bf{\alpha}}_j+\partial_{\beta_j})
+  \frac{1}{2} U\sum_{i=1}^M \beta_i^2{({\bf{\alpha}}_i+\partial_{\beta_i})}^2\nonumber\\
&-&\mu_e \sum_{i=1}^M
\beta_i ({\bf{\alpha}}_i+\partial_{\beta_i})
\end{eqnarray}
\begin{figure}
\hspace{-0.3cm}
\begin{center}
$\begin{array}{c}
\includegraphics[angle=0,width=7.3cm,height=5cm]{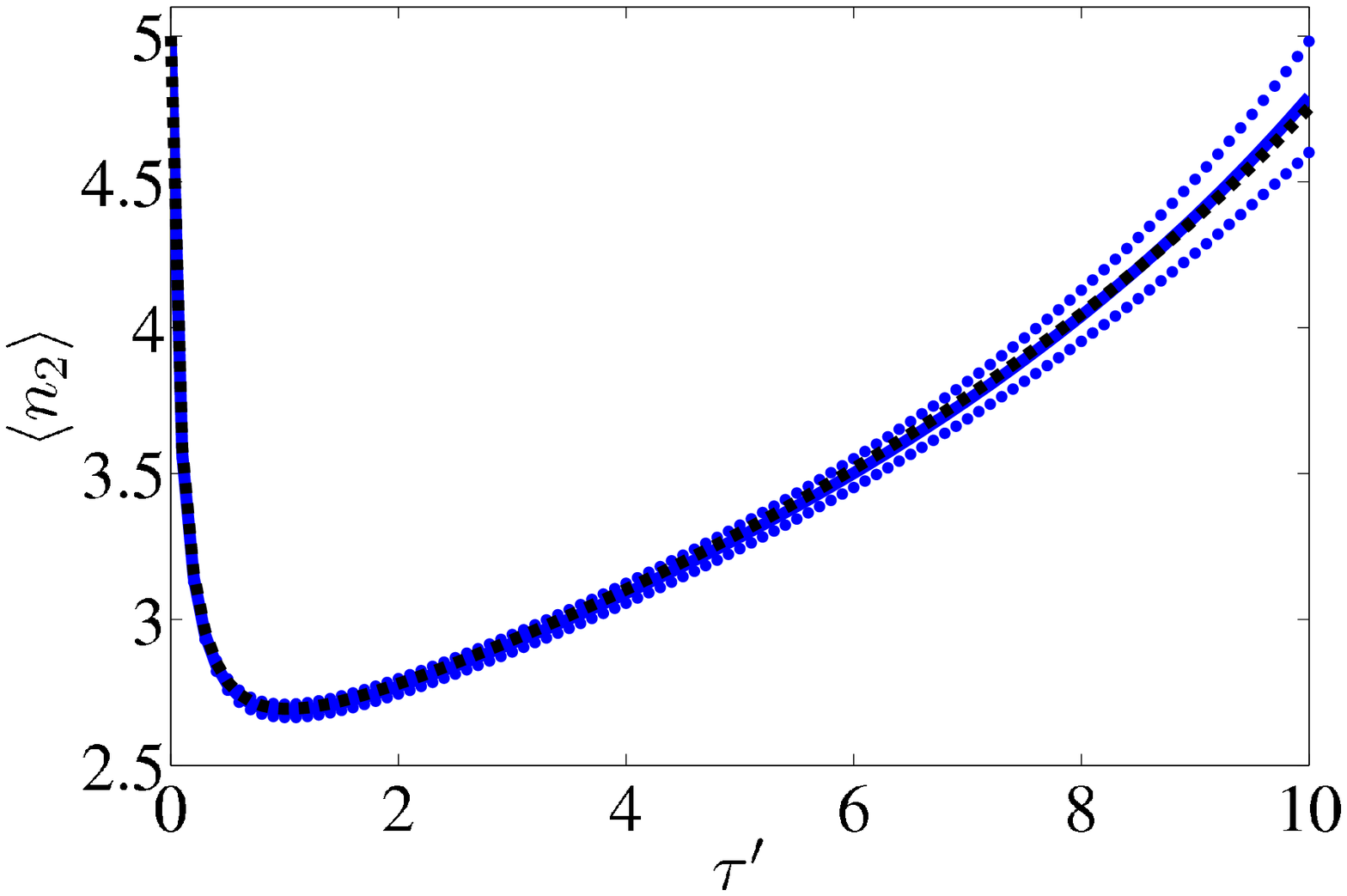}
\end{array}$
\caption[Comparison of the simulation and analytical results for
$M=2$ and $U=0$.]{\small Comparison of the simulation results for
 $\bra{\hat n}_2\ket$ with the analytical
results, for $n_0=5$, $M=2$, $U=0$, $J=1$ and $\mu_{{}_T}/J=-1.01$
($\mu_e/J=-0.9918$).  Here, we have
$\tau^\prime={J/{k_{B}T}}={1/{T}}$ and the number of the stochastic
trajectories $n_p$ is $10^5$.  Other conventions as in
\fref{M1nos_Exact_and_Numerical}.}\label{M2U0n1n2}
\end{center}
\end{figure}
\begin{figure}%[t]
\begin{center}
$\begin{array}{c}
\includegraphics[angle=0,width=8.cm,height=5cm]{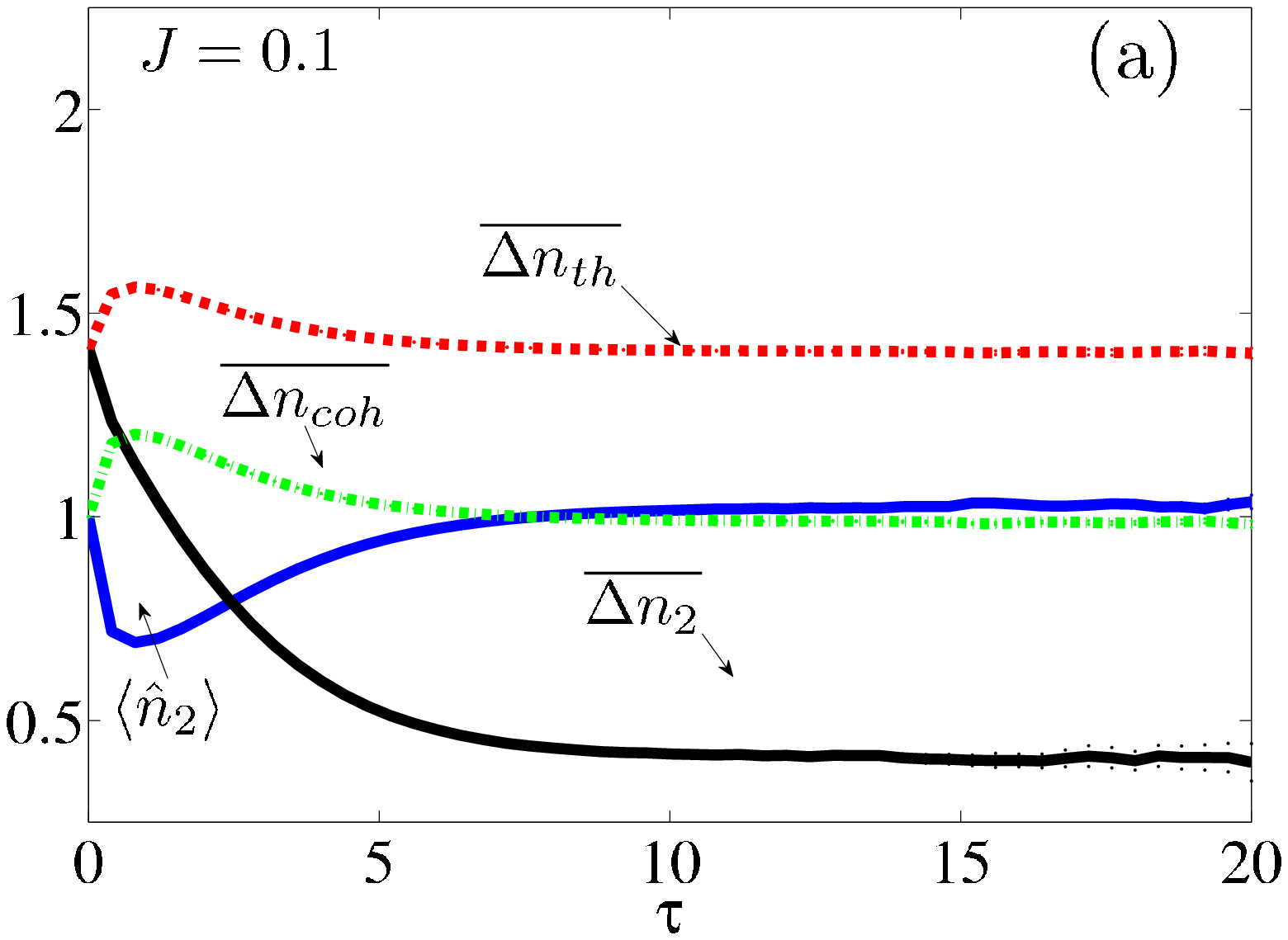}\\
\includegraphics[angle=0,width=8.cm,height=5cm]{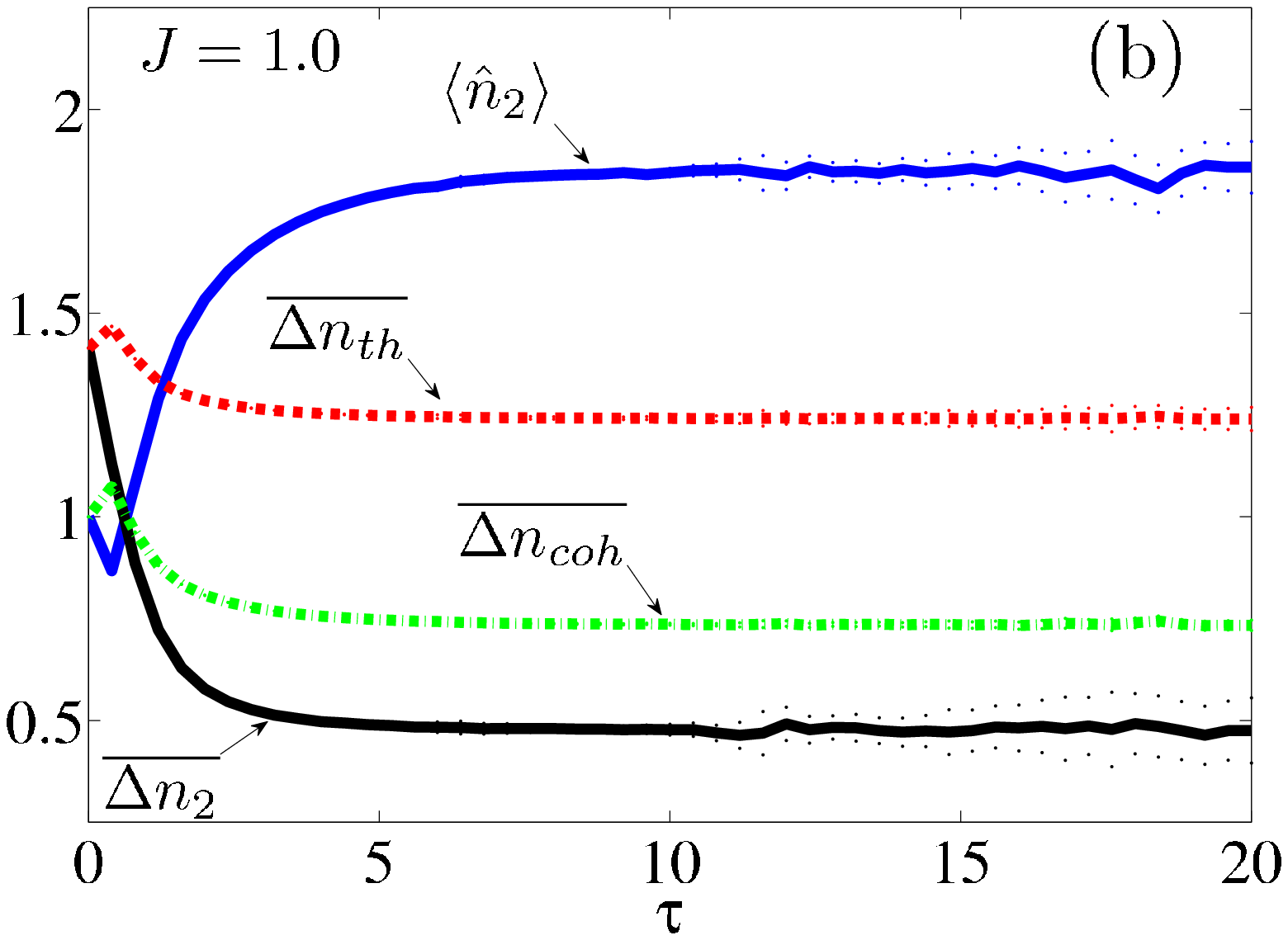}
\end{array}$
\caption[$\bra{\hat n}_2\ket$, $\overline{\Delta{n_2}}$,
$\overline{\Delta {n_{th}}}$ and $\overline{Delta{n_{th}}}$ versus
the $\mu_{{}_T}$ for a double well.]{\small Simulation results for
the expectation values of the number of particles $\bra{\hat
n}_2\ket$, the relative standard deviation
$\overline{\Delta{n_2}}=\Delta{n_2}/\bra{\hat n}_2\ket$, the
relative standard deviation for a coherent state with the same
number of particles
$\overline{\Delta{n_{coh}}}=\Delta{n_{coh}}/\bra{\hat n}_2\ket$ and
the relative standard deviation for a thermal state with the same
number of particles
$\overline{\Delta{n_{th}}}=\Delta{n_{th}}/\bra{\hat n}_2\ket$ at one
of the sites in a double well system. $\tau$ is proportional to the
inverse temperature and the dotted lines around each line show the
sampling errors. Here, the stochastic averages have been taken over
$n_p=10^9$ trajectories and the target chemical potential
$\mu_{{}_T}$ is $0.5$.}\label{M2n1U1Jall}
\end{center}
\end{figure}
\subsection{Two-site model for $U=0$}\label{Two-site model}
For $M=2$, we have a double well and the  Stratonovich equations can be written from \eref{StraOmega} and~\eref{StraAlphaj}.
\Fref{M2U0n1n2} shows $\bra{\hat n}_2\ket$
for a double site model ($M=2$) with $U=0$, $J=1$, $n_0=5$ and
$\mu_{{}_T}/J=-1.01$ ($\mu_e/J=-0.9918$), where $\mu_{{}_T}=-1.01$
is the target chemical potential at $T=0.1{\hspace{0.1cm}}J$. This
figure also compares the exact analytical results~\cite{GhanbariThesis} with the gauge $P$ simulations and shows good
agreement between them.  The sampling error is due to the stochastic
initial conditions given by~\eref{initial}.
\subsection{Two-site model for $U\neq0$ and $J\neq0$}
The simulation codes have been tested for the limiting cases when either $U=0$ or $J\neq0$
 and are now ready for the general case of a
two-site system in which both hopping matrix element $J$ and the
on-site interaction  $U$ exist.
\begin{figure*}
\begin{center}
$\begin{array}{ccc}
\includegraphics[angle=0,width=5.7cm]{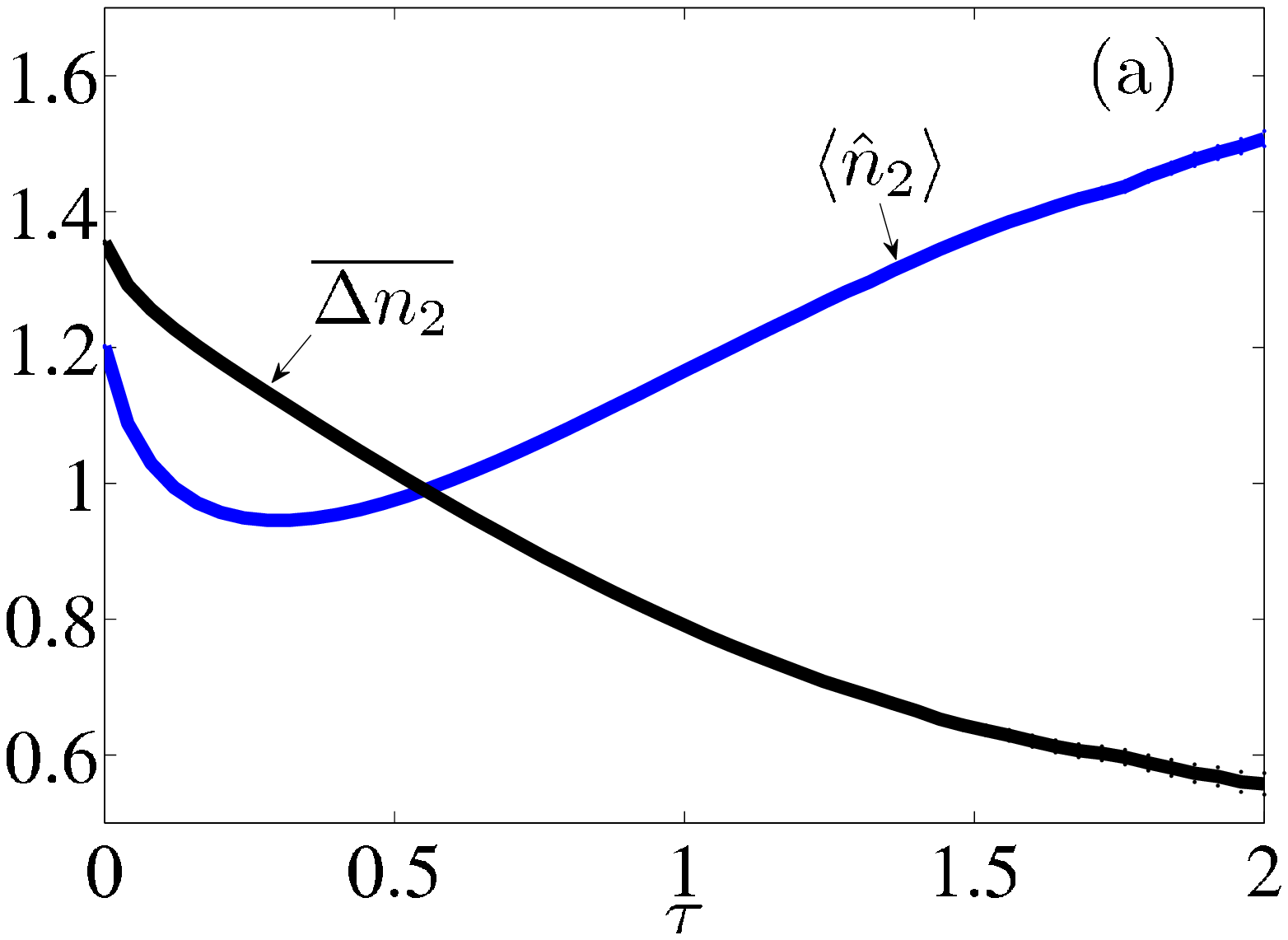}
\includegraphics[angle=0,width=5.7cm]{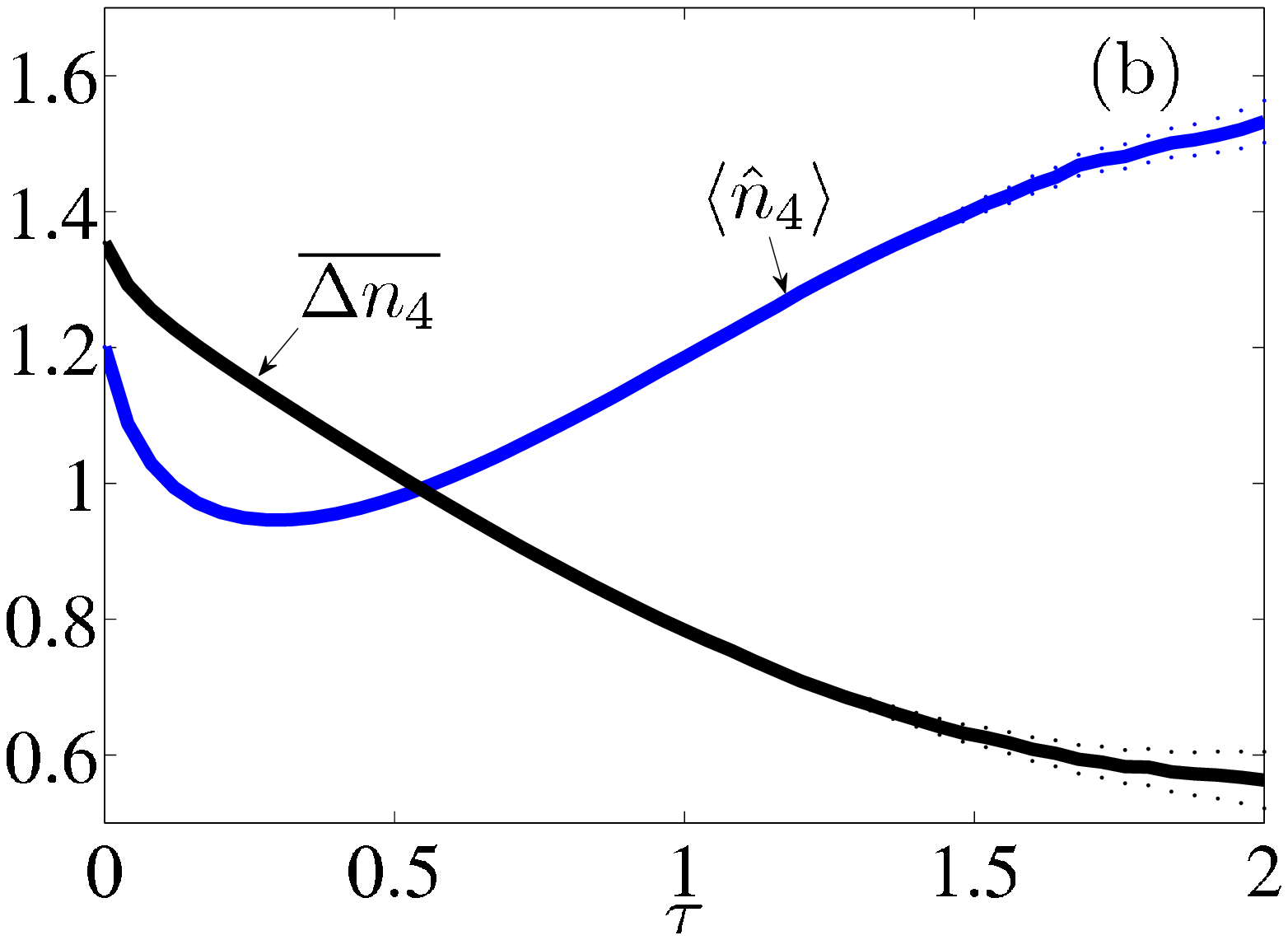}
\includegraphics[angle=0,width=5.7cm]{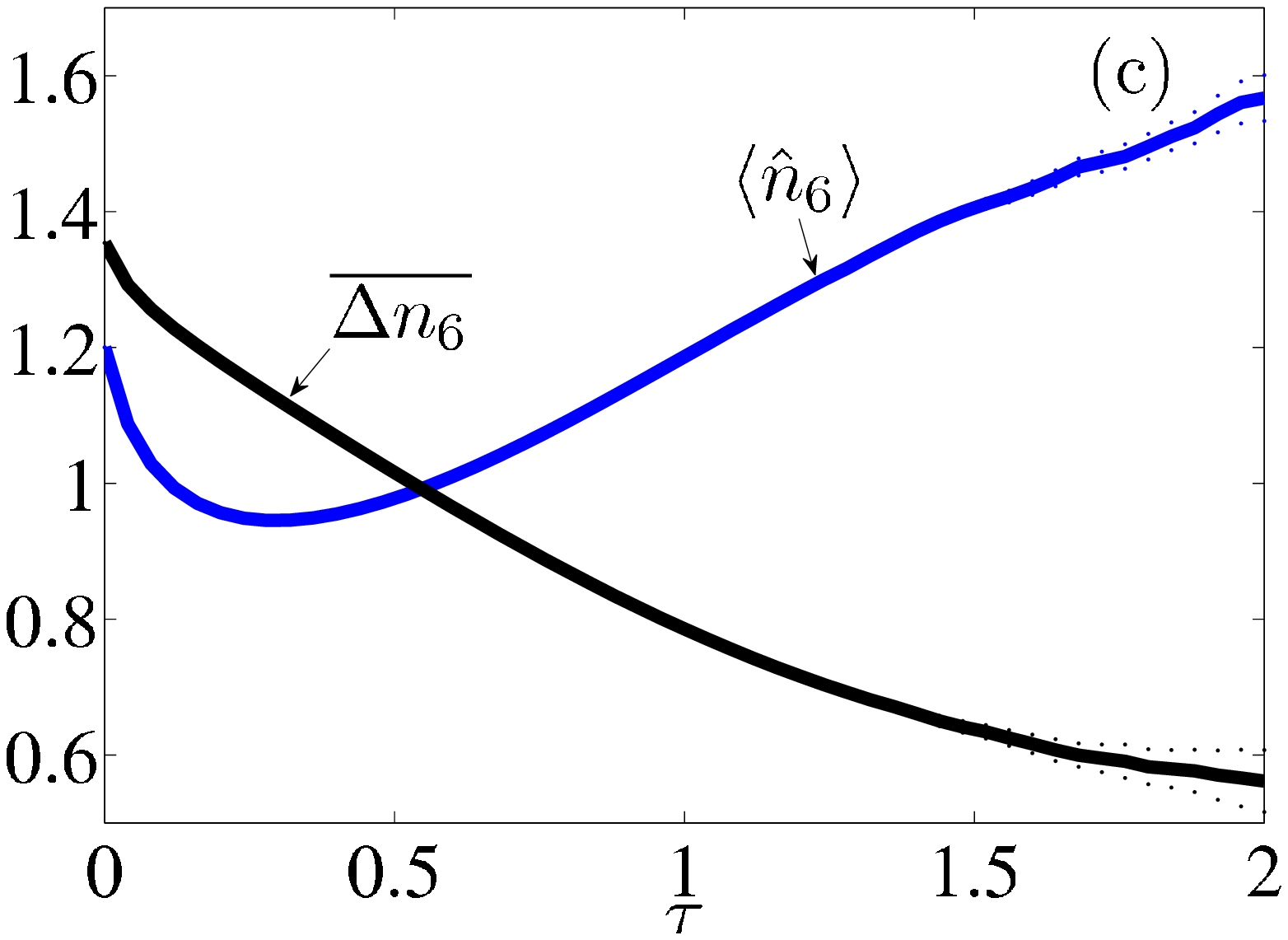}
\end{array}$
\caption[Size effect]{\small $\langle{\hat{n}}\rangle$ and the relative standard
deviation $\overline{\Delta n}=\Delta n/\langle{\hat n}\rangle$ for the hopping matrix
element $J=0.4$ and three different system sizes $M=3$,
$7$ and $11$. Here, we also have $n_0=1.2$, $U=1.0$ and
$\mu_{{}_T}=0.5$. When the size of the system changes from 7 to 11,
for the central sites, 4 and 6, respectively, the expectation values
of the number of particles and especially the standard deviations, within
the sampling error, are in good
agreement.}\label{Deltan_n_Tau2_M3_7_11J0__0_4}
\end{center}
\end{figure*}
\begin{figure*}%[h]
\begin{center}
$\begin{array}{cc}
\includegraphics[angle=0,width=7.7cm,height=4.5cm]{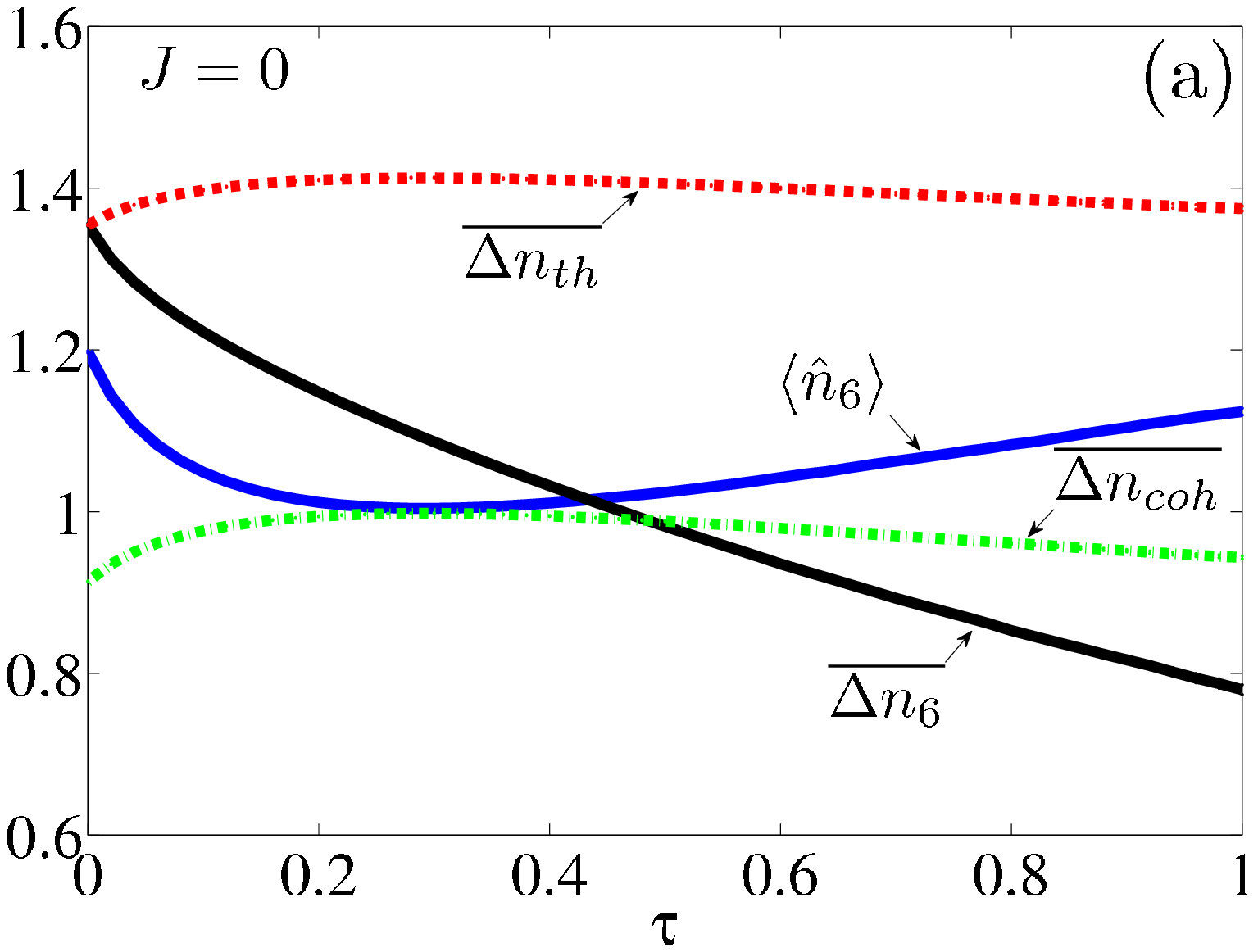}
\includegraphics[angle=0,width=7.7cm,height=4.5cm]{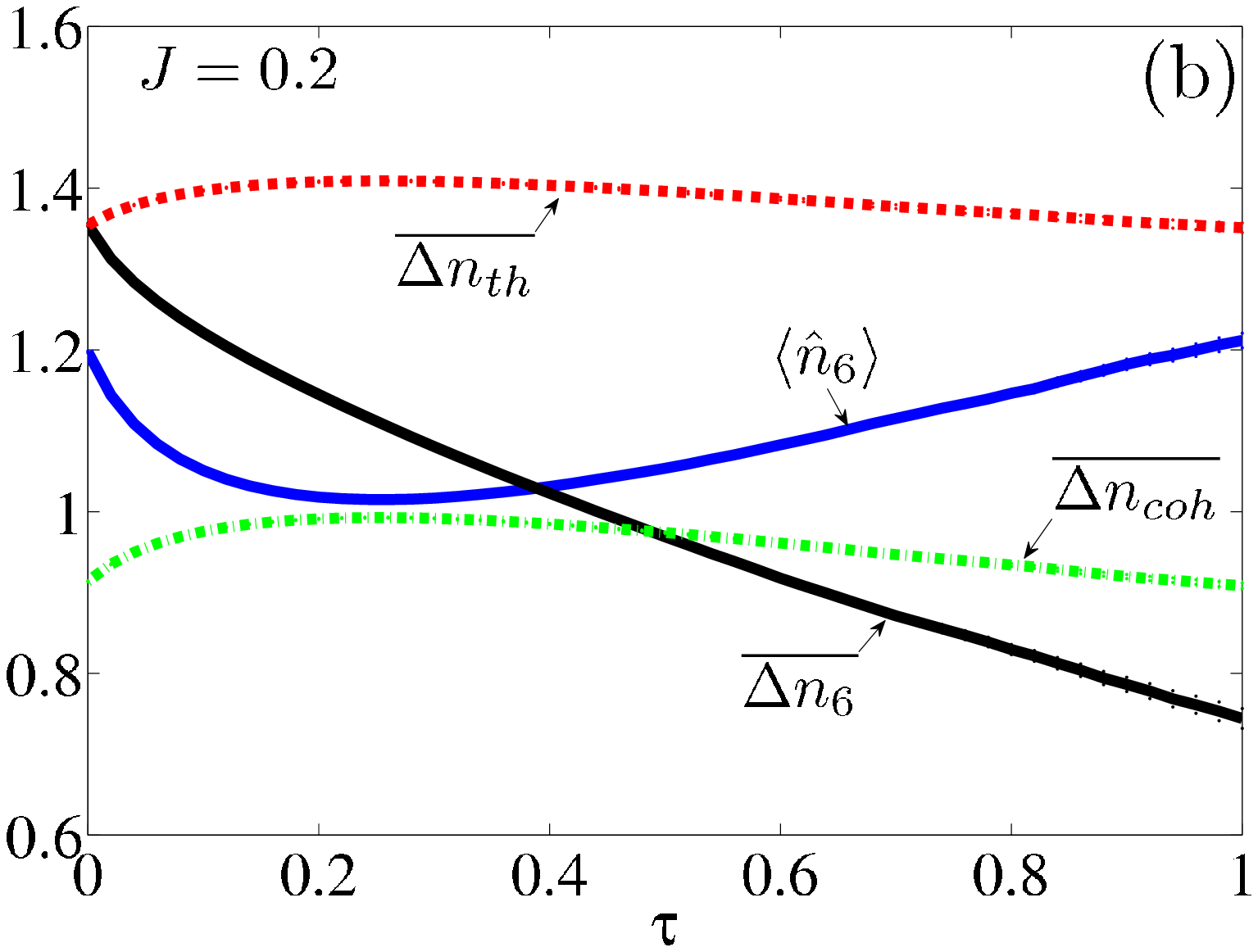}\\
\includegraphics[angle=0,width=7.7cm,height=4.5cm]{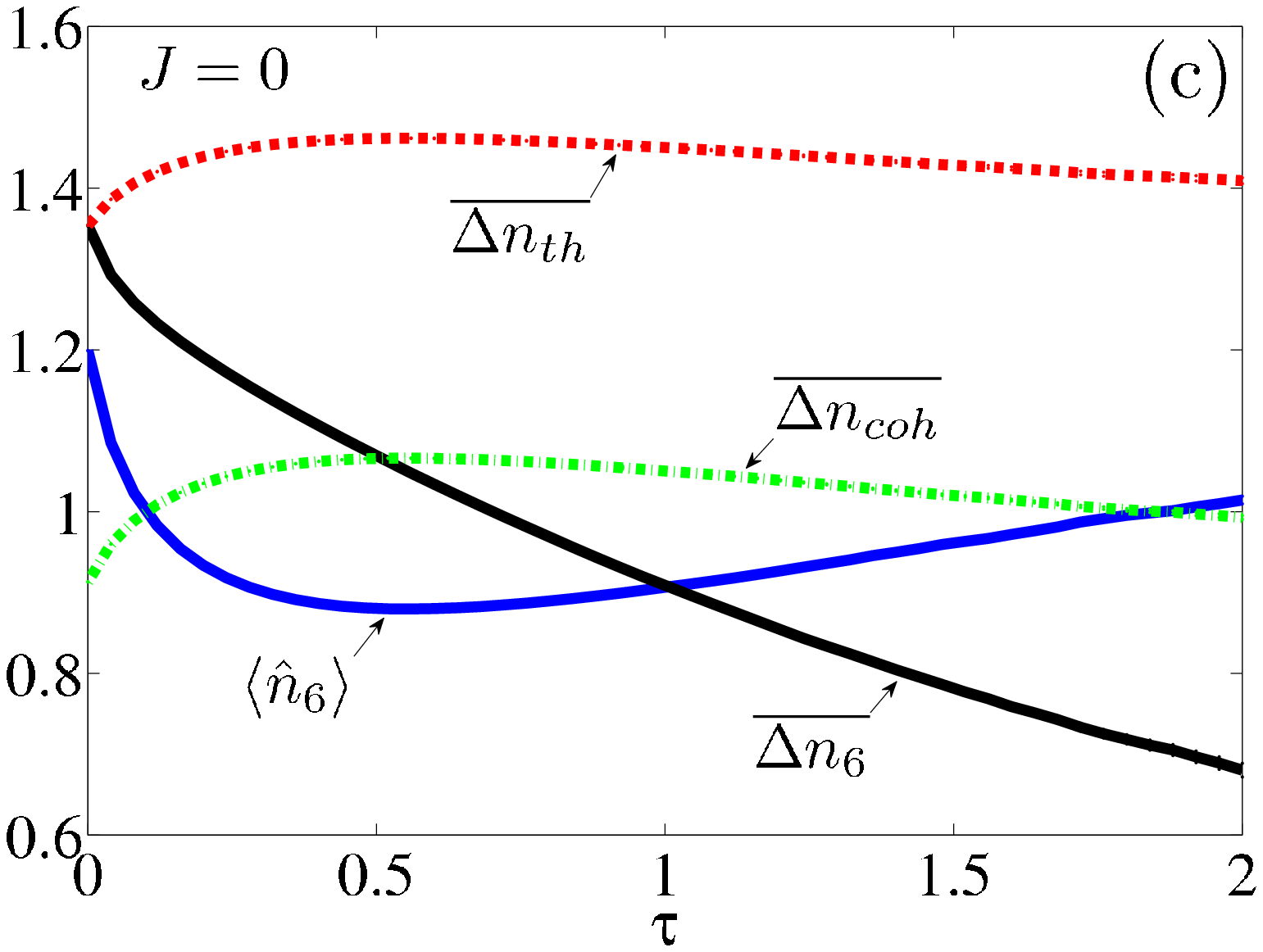}
\includegraphics[angle=0,width=7.7cm,height=4.5cm]{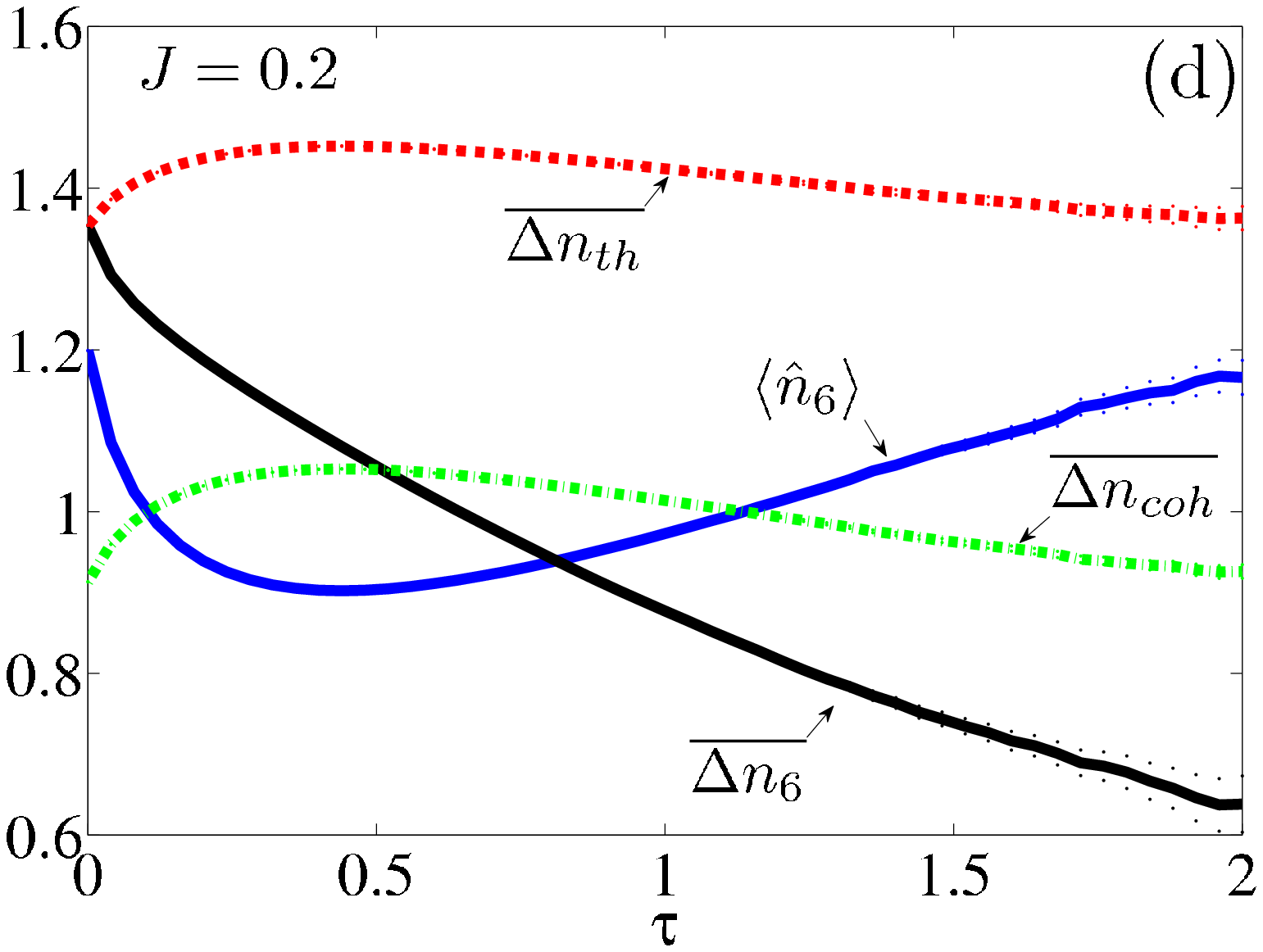}
\end{array}$
\caption[Standard deviations
 for $M=11$, $n_0=1.2$, $U=1.0$,
$\mu_{{}_T}=0.5$]{\small $\langle{\hat n}_6\rangle$ and the relative standard
deviations, for a general state $\overline{\Delta{n_6}}$, a thermal state
$\overline{{\Delta n}_{th}}$ and  the coherent state $\overline{{\Delta n}_{coh}}$. Here
we have $M=11$, $n_0=1.2$, $U=1.0$, $\mu_{{}_T}=0.5$ and (a) $J=0$,
$\tau_{{}_T}=1$, $n_p=10^6$, (b) $J=0.2$, $\tau_{{}_T}=1$,
$n_p=10^6$ , (c) $J=0$, $\tau_{{}_T}=2$, $n_p=10^7$  and (d)
$J=0.2$, $\tau_{{}_T}=2$, $n_p=10^7$.
}\label{M11Deltan6muT0_5J_0__0_2tau1_2}
\end{center}
\end{figure*}
In \fref{M2n1U1Jall} $\bra{\hat n}_2\ket$ (blue solid line), the
relative standard deviation
$\overline{\Delta{n_2}}=\Delta{n_2}/\bra{\hat n}_2\ket$ (black solid
line), the relative standard deviation for a coherent state with the
same number of particles
$\overline{\Delta{n_{coh}}}=\Delta{n_{coh}}/\bra{\hat n}_2\ket$
(green dashed-dotted line)  and the relative standard deviation for
a thermal state with the same number of particles
$\overline{\Delta{n_{th}}}=\Delta{n_{th}}/\bra{\hat n}_2\ket$ (red
dashed line) in a double well system are shown, where $\Delta{n_2} =
\sqrt{\bra{\hat n}^2_2\ket-\bra{\hat n}_2\ket^2}$, $\Delta{n_{coh}}
= \sqrt{\bra{\hat n}_2\ket}$ and $\Delta{n_{th}} = \sqrt{\bra{\hat
n}_2\ket^2+\bra{\hat n}_2\ket}$. Here the on-site interaction $U$ is
$1$ and $J/U$ in \fref{M2n1U1Jall}(a) and (b) is $0.1$ and $1.0$,
respectively. In this quantum simulation, the number of
trajectories, $n_p$,  is $10^9$.  Also, the target chemical
potential, $\mu_{{}_T}$,  is $0.5$.

As \fref{M2n1U1Jall}(a) shows, when $J/U$ is $0.1$, the average
number of particles at the temperature $T=0.05\hspace{0.1cm}U$
($\tau_{{}_T}=20$) is almost 1 which is close to the exact value of
the average number of particles for $J/U=0$, shown in  \fref{M1_n1_tau_20_j0_muT_0_5}.
  Note that,
as we increase $J/U$ to $1.0$,
at $\tau_{{}_T}=20$, we have $\bra{\hat n}_2\ket\simeq 1.8$.  Also,
if we increase the hopping matrix element the relative standard deviation $\overline{\Delta{n_2}}$
increases and in both cases for $J/U=0.1$ and $J/U=1.0$ we have
\begin{equation}\label{fluctuations}
\overline{\Delta{n_2}} < \overline{\Delta{n_{coh}}} <\overline{\Delta{n_{th}}}
\end{equation}
It is interesting also that the relative standard deviation $\overline{\Delta{n_2}}$
approaches that of a coherent state as we move to large values of
$J/U$ which at low temperatures are in the superfluid regime.  This
is in the right direction for describing a superfluid as a coherent
state.
\section{Simulation of M-site model for $U\neq0$ and $J\neq0$}\label{M-site model}
For $M\geq 3$, the Stratonovich equations can be written from \eref{StraOmega} and \eref{StraAlphaj}.
\begin{figure*}%[t]
\begin{center}
$\begin{array}{ccc}
\includegraphics[angle=0,width=5.8cm,height=4.0cm]{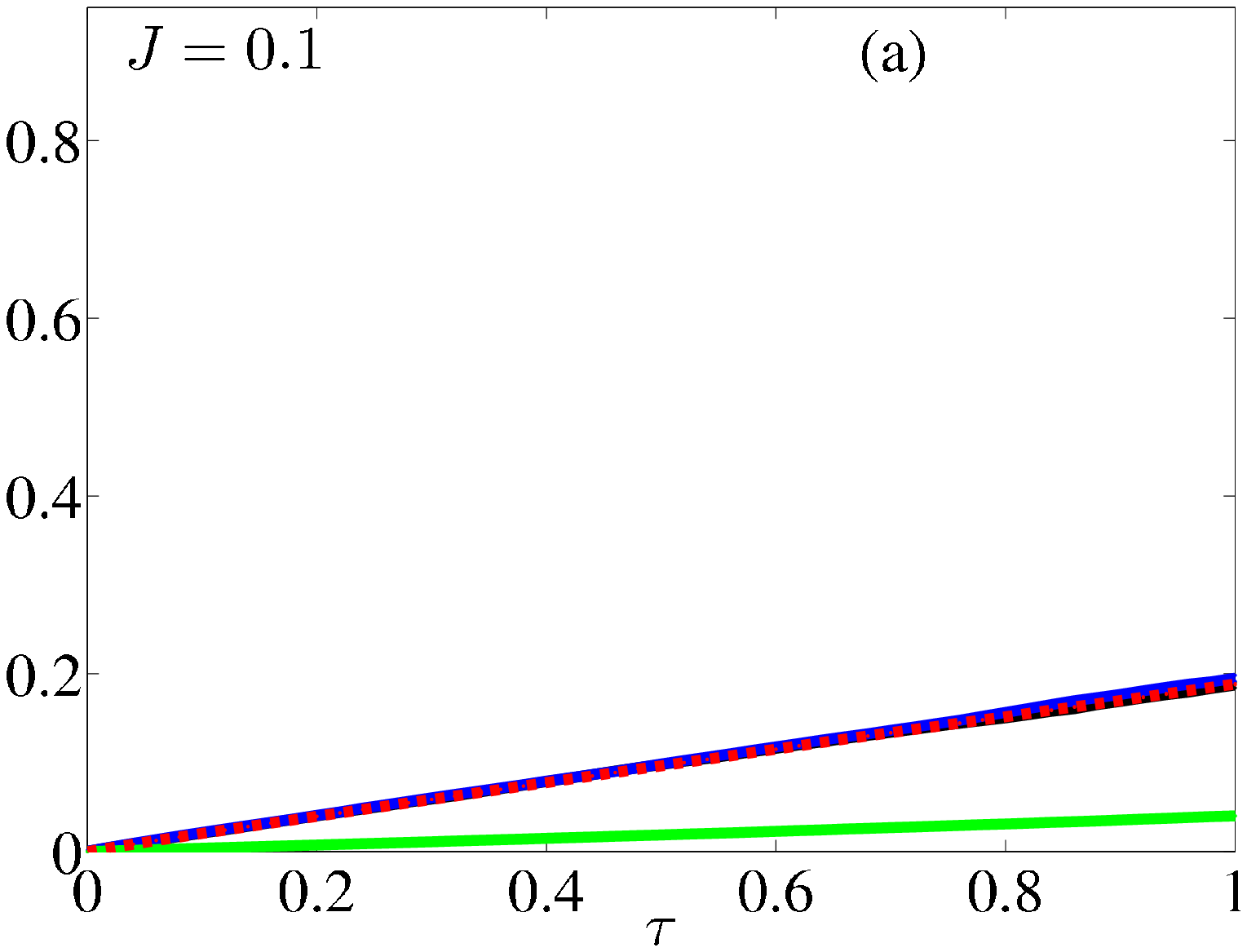}
\includegraphics[angle=0,width=5.8cm,height=4.0cm]{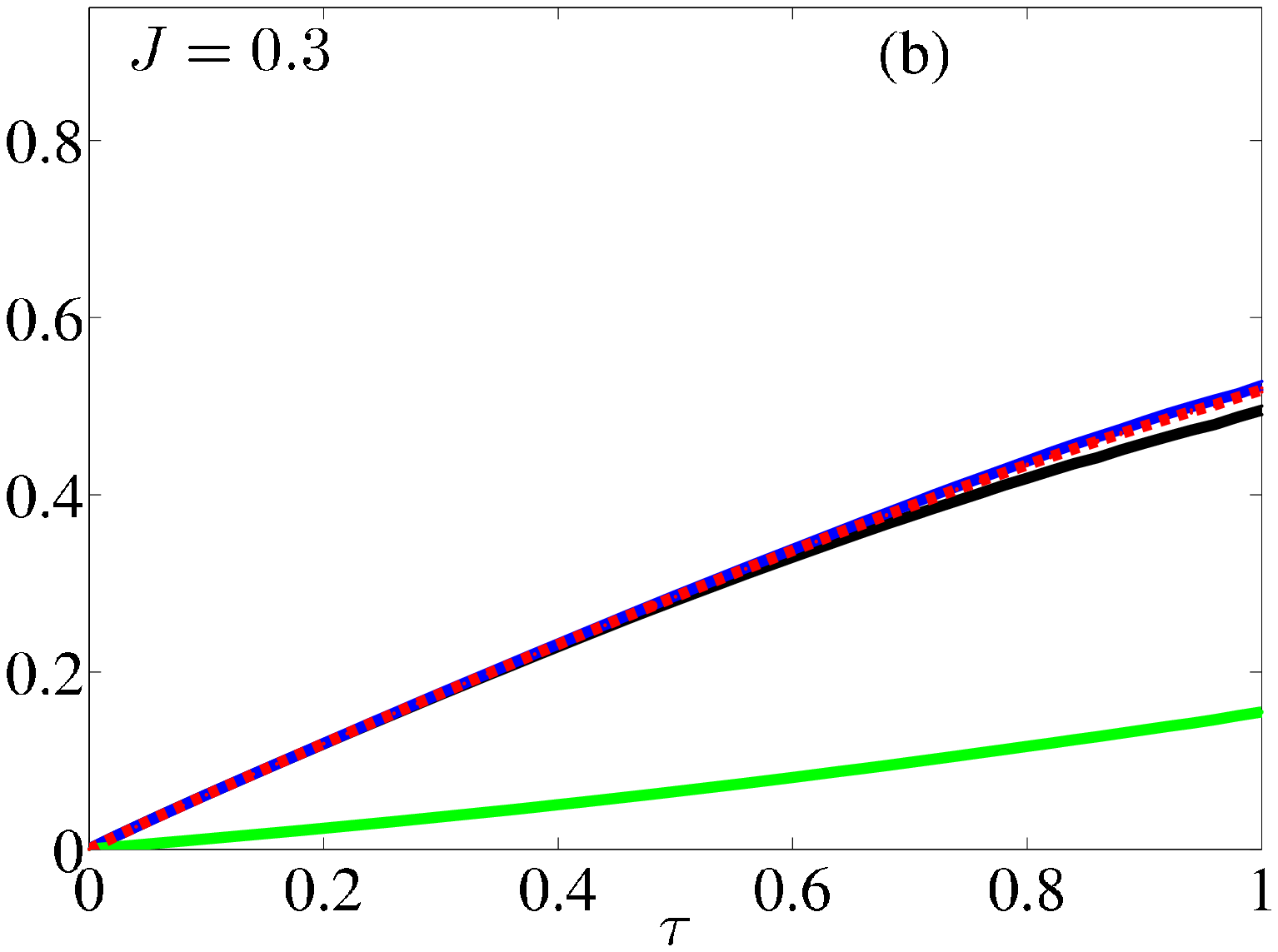}
\includegraphics[angle=0,width=5.8cm,height=4.0cm]{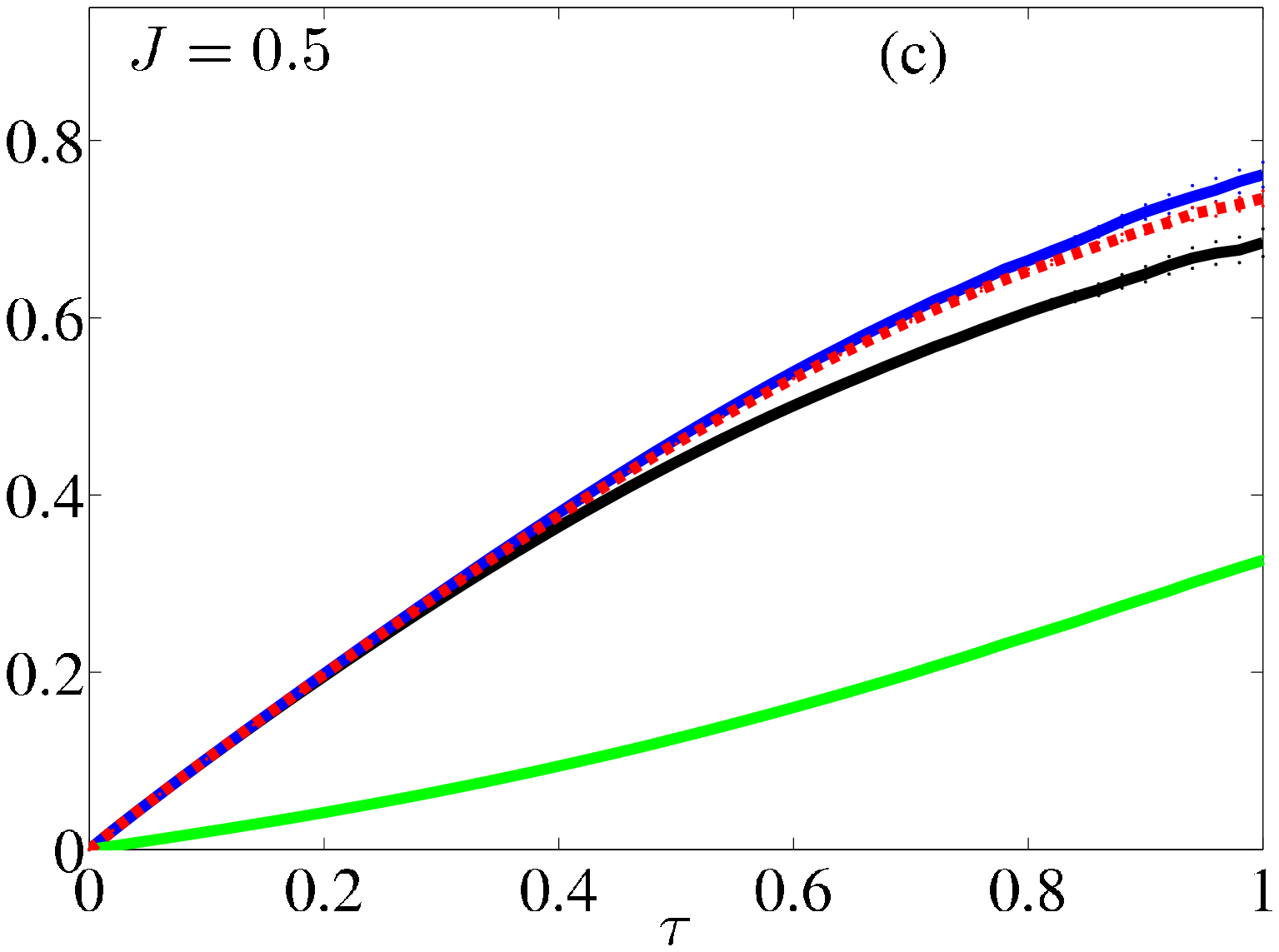}\\
\includegraphics[angle=0,width=5.8cm,height=4.0cm]{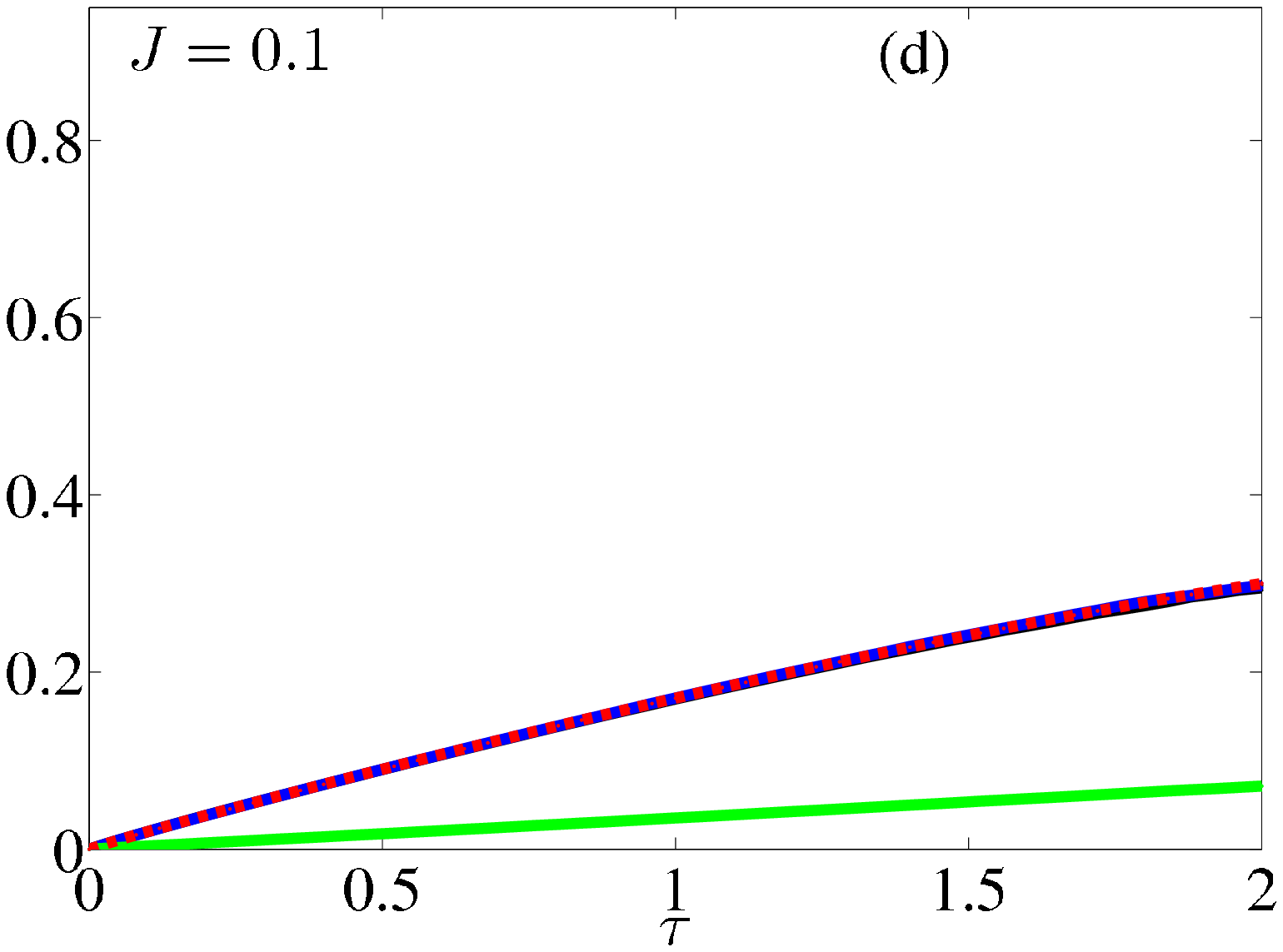}
\includegraphics[angle=0,width=5.8cm,height=4.0cm]{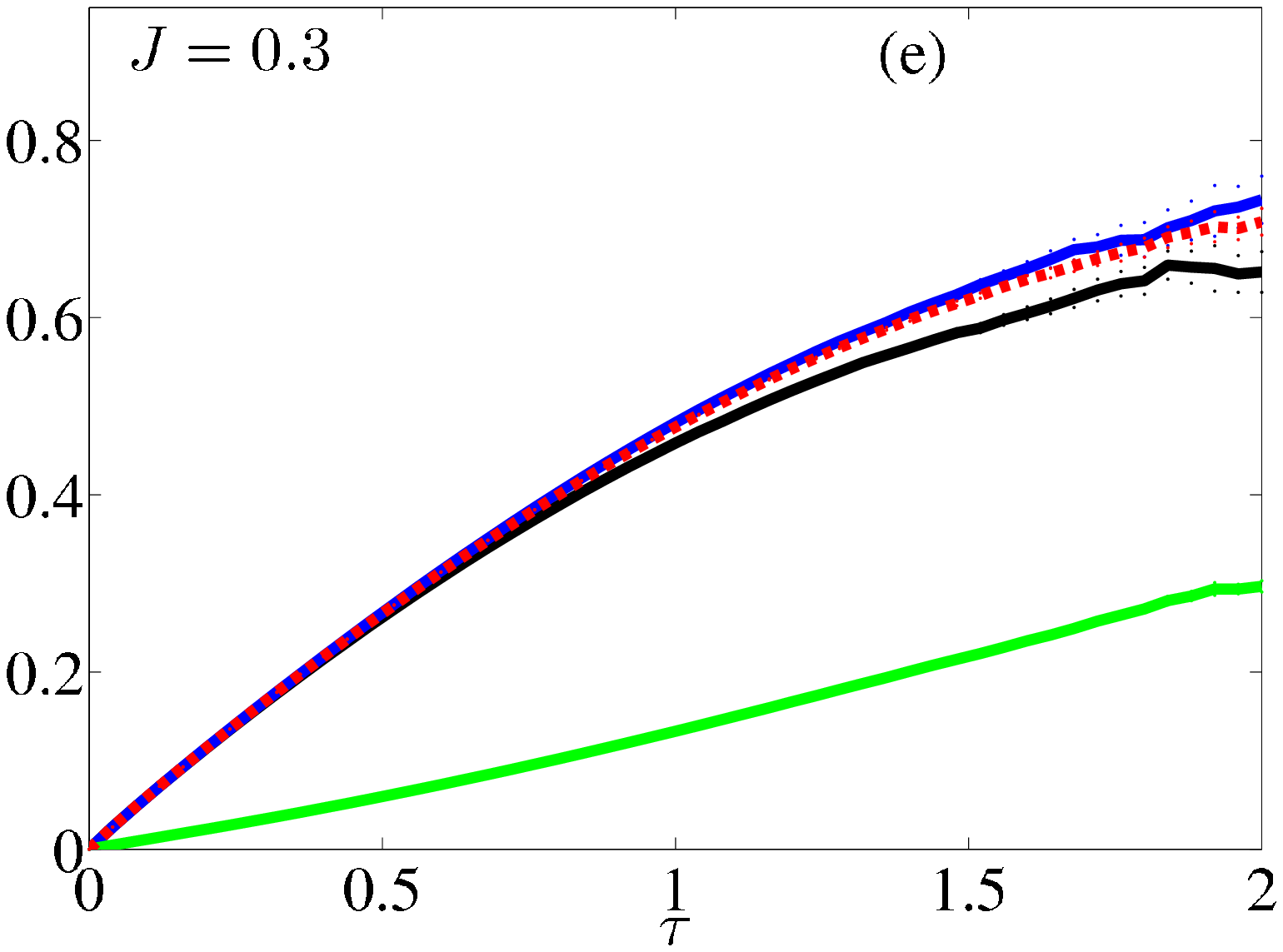}
\includegraphics[angle=0,width=5.8cm,height=4.0cm]{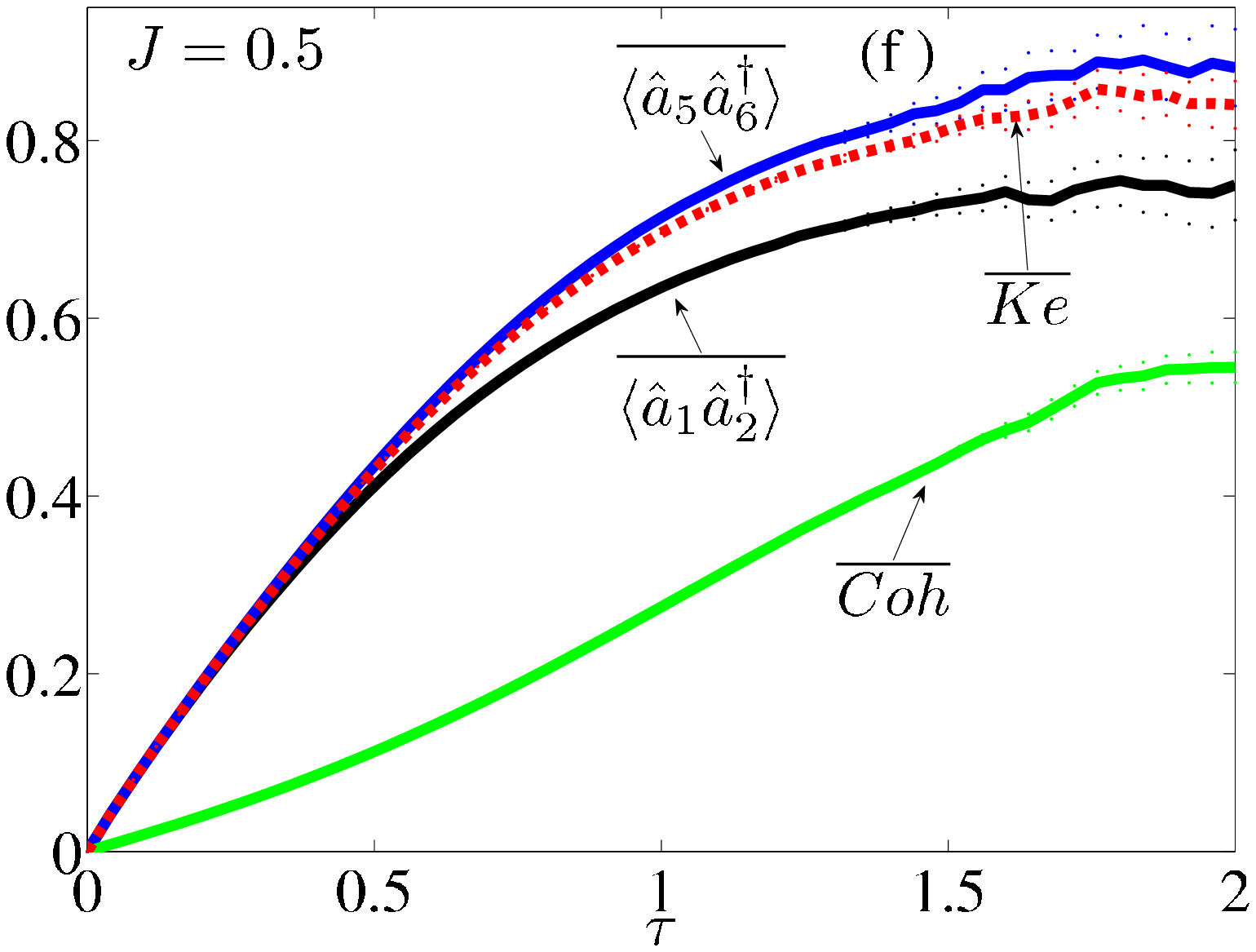}
\end{array}$
\caption[Relative values of the coherence between sites.]{\small
This figure shows relative values of the coherence between sites,
$\overline{Coh}=Coh/\bra {\hat n}_6\ket$, $\overline{Ke}=Ke/\bra {\hat n}_6\ket$, $\overline{\bra {\hat a}_1 {\hat a}_{2}^\dagger\ket}=\bra {\hat a}_1 {\hat a}_{2}^\dagger\ket/\bra {\hat n}_6\ket$ and
 $\overline{\bra {\hat a}_5 {\hat a}_{6}^\dagger\ket}=\bra {\hat a}_5 {\hat a}_{6}^\dagger\ket/\bra {\hat n}_6\ket$ at two different
 target temperatures  $T=\hspace{0.1cm}U$ in (a) and (c) and $T=0.5\hspace{0.1cm}U$ in (b) and (d).
  The labels in (a)-(c) are the same as in (d).  By decreasing the target temperature
  (increasing $\tau_{{}_T}$) $\overline{Coh}$, $\overline{Ke}$,
  $\overline{\bra {\hat a}_1 {\hat a}_{2}^\dagger\ket}$ and
 $\overline{\bra {\hat a}_5 {\hat a}_{6}^\dagger\ket}$ increase.
 Also, at a constant temperature, by increasing the hopping matrix element,  all of
 these measures of the coherence between lattice sites increase.}\label{M11coh_ke_a_adags_tau_j_Rel}
\end{center}
\end{figure*}
\Fref{Deltan_n_Tau2_M3_7_11J0__0_4} compares $\langle{\hat{n}}\rangle$ and
$\overline{\Delta n}={\Delta n}/\langle{\hat{n}}\rangle$ at the central
site for three different system sizes with the hopping matrix element $J=0.4$.
This  figure shows that the size effect is not very considerable for
$M=11$ as, within the sampling error, the values of $\langle{\hat
n}\rangle$ and $\overline{\Delta n}$  are similar to those for
$M=7$ and, in particular the relative number fluctuations are the same.  Therefore, a Bose-Hubbard model with $M=11$ may be large
enough to approximately represent the behaviour of an infinite system.
\begin{figure*}%[th]
\begin{center}
$\begin{array}{cc}
\includegraphics[angle=0,width=8.cm,height=4.5cm]{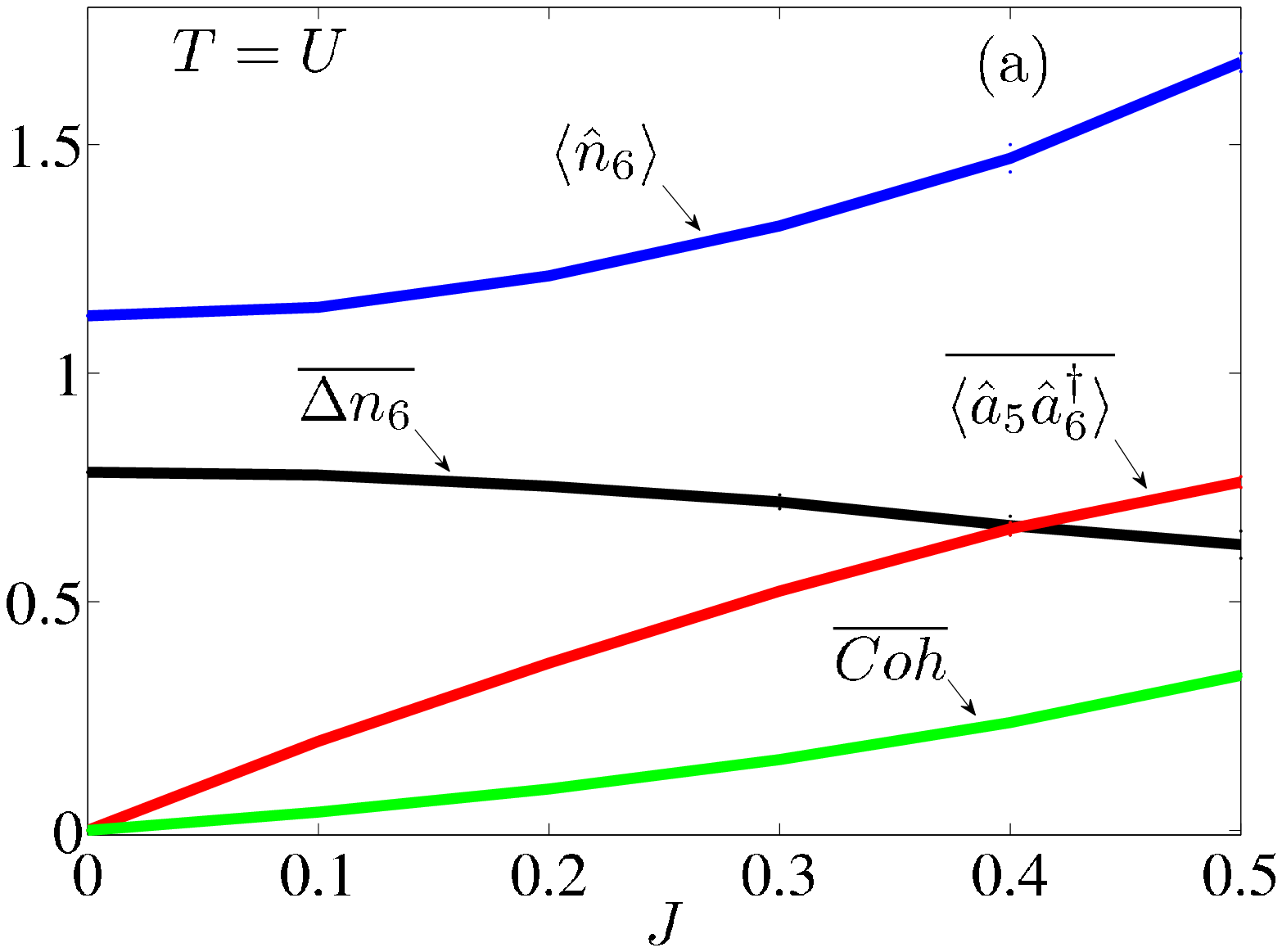}
\includegraphics[angle=0,width=8.cm,height=4.5cm]{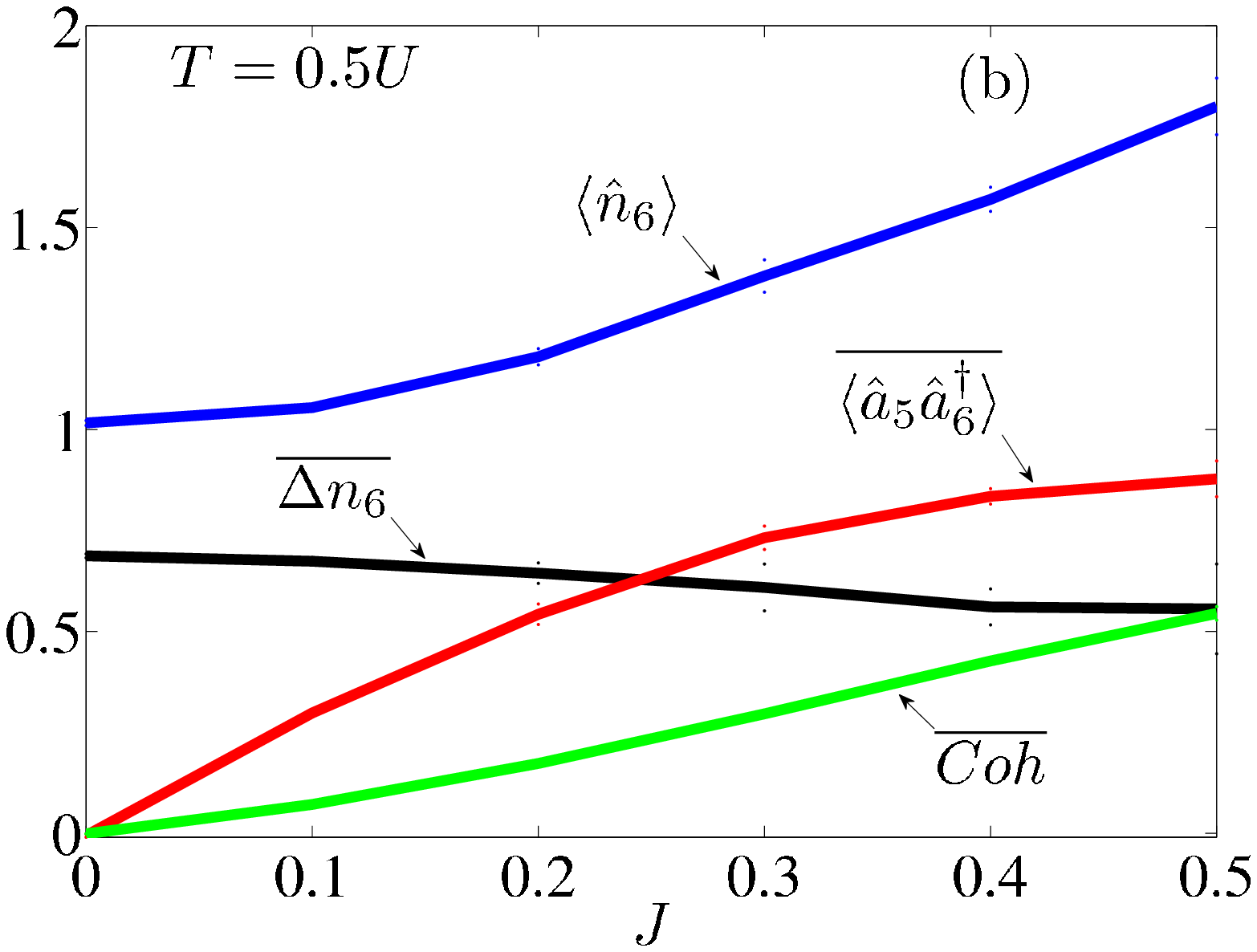}\\
\includegraphics[angle=0,width=8.cm,height=4.5cm]{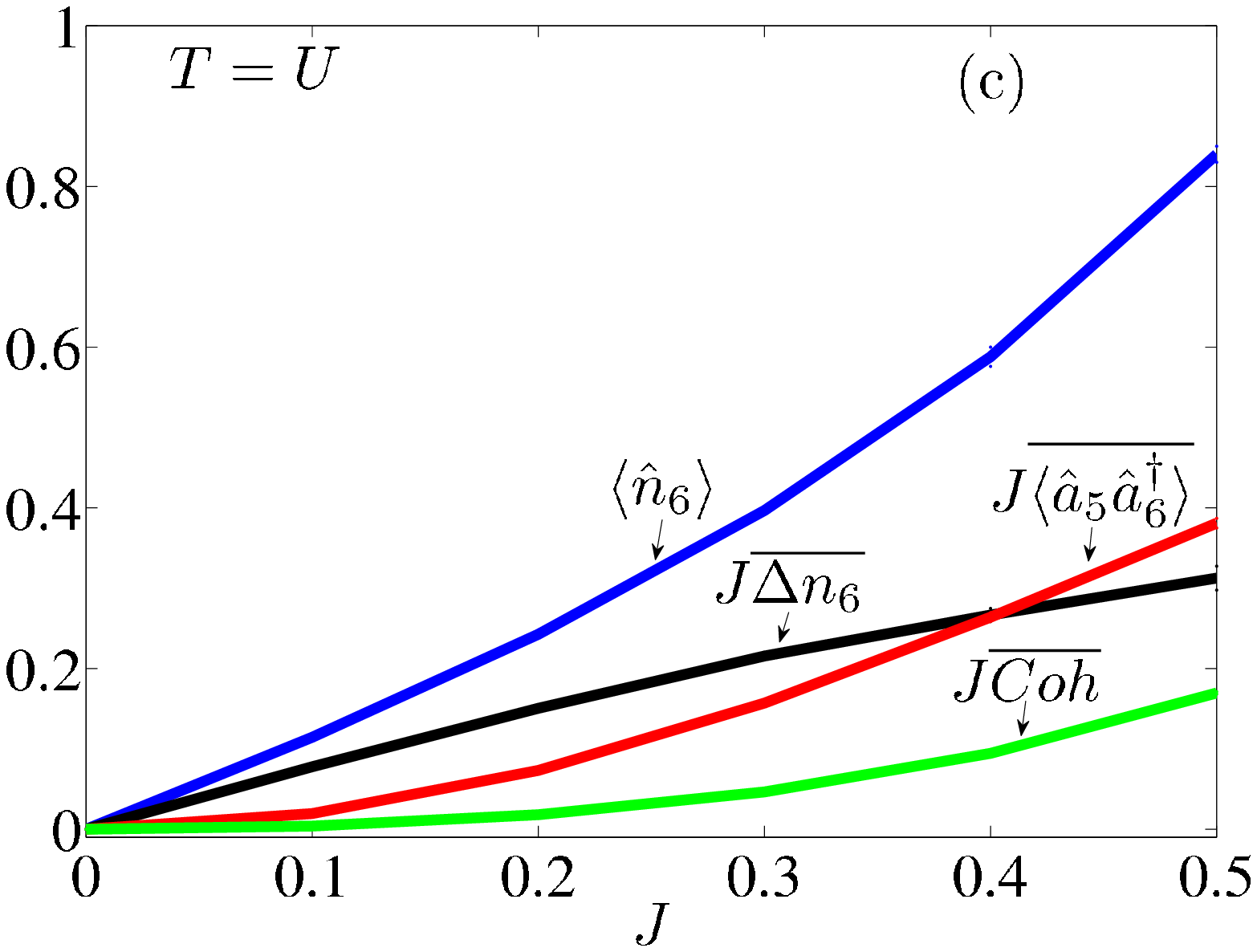}
\includegraphics[angle=0,width=8.cm,height=4.5cm]{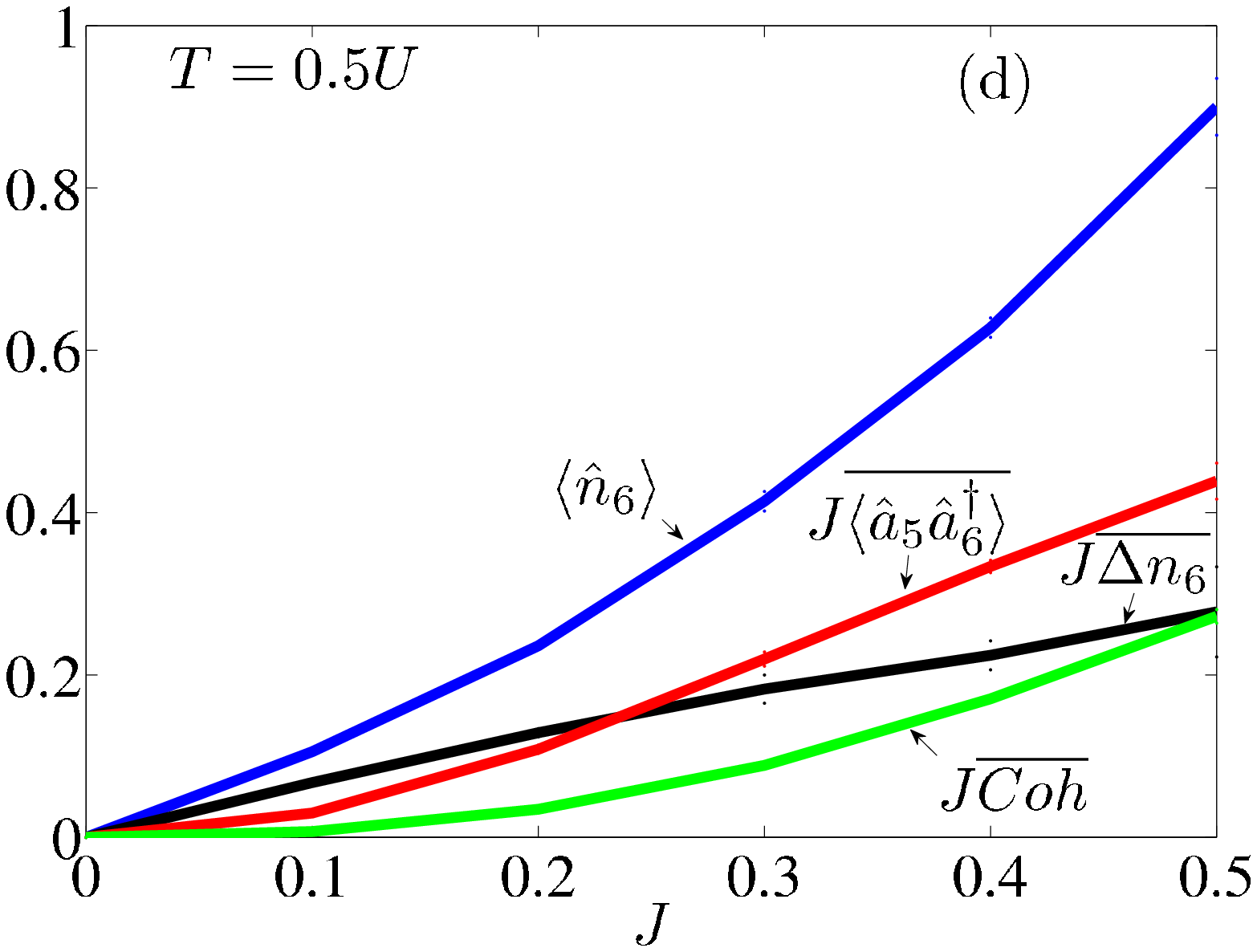}
\end{array}$
\caption[$\bra{\hat n}_6 \ket$ and the relative values $\overline{\Delta{n_6}}$, $\overline{\bra {\hat a}_5
{\hat a}_6^\dag\ket}$, $\overline{Coh}$, $J\overline{\bra{\hat n}_6 \ket}$, $J\overline{\Delta{n_6}}$,
$J\overline{\bra {\hat a}_5 {\hat a}_6^\dag\ket}$ and  $J\overline{ Coh}$.]{\small
${\bra{\hat n}_6 \ket}$ and the relative values $\overline{\Delta{n_6}}$, $\overline{\bra {\hat a}_5 {\hat
a}_6^\dag\ket}$, $\overline{Coh}$, $J\overline{\bra{\hat n}_6 \ket}$, $J\overline{\Delta{n_6}}$,
$J\overline{\bra {\hat a}_5 {\hat a}_6^\dag\ket}$ and  $J \overline{Coh}$ as a function of
$J$ for two different target temperatures $T=0.5 U$ and $T=U$. All
lines are to guide the eye. The sampling error, which is small, is
shown by the dots along the lines.}\label{coh_a5a6dag_n_delta_n}
\end{center}
\end{figure*}
In \fref{M11Deltan6muT0_5J_0__0_2tau1_2} the relative standard deviations $\overline{\Delta n_6}$,
, are plotted versus the inverse temperature, $\tau$,  for
a general state and are compared with those of an ideal thermal state $\overline{\Delta{n_{th}}}$, where
$\left(\Delta n_{th} = \sqrt {{\bra \hat{n} \ket}^2+\bra \hat{n} \ket}\right)$, and a
coherent state $\overline{\Delta{n_{coh}}}$, where $\left(\Delta n_{coh} = \sqrt {\bra \hat{n} \ket}\right)$ for two different values of $J/U$.  In order to obtain $\Delta n$ for the thermal and the
coherent states we have used the simulation results for $\bra \hat{n}
\ket$.
According to the plots, at  a constant low temperature, as $J/U$ is
increased, the standard deviation increases.
 Also the standard deviation at the target temperature $T$ and the target
 chemical potential $\mu_{{}_T}$for both different values of  $J/U$ is  less
 than the standard deviation for  a coherent state and much less than that of
 a thermal state. Also because at low values of $J/U$ the atoms form a superfluid,
 when the temperature goes to zero,
 the closeness of the standard deviation $\overline{\Delta n_6}$ to $\overline{{\Delta n}_{coh}}$
 supports the idea of describing a superfluid by a coherent state.
 Also an increase in the (quantum) standard deviation $\bra n \ket$
 is observed as the temperature increases.

Because the number of atoms changes with temperature, it is
important to know the behaviour of the measures of the coherence
relative to the average number of atoms per lattice site.
\Fref{M11coh_ke_a_adags_tau_j_Rel} shows relative coherences
$\overline{\bra {\hat a}_1 {\hat a}_{2}^\dagger\ket}=\bra {\hat a}_1
{\hat a}_{2}^\dagger\ket/\bra{\hat n}_6\ket$ and $\overline{\bra
{\hat a}_5 {\hat a}_{6}^\dagger\ket}=\bra {\hat a}_5 {\hat
a}_{6}^\dagger\ket/\bra{\hat n}_6\ket$ and also
$\overline{Coh}=Coh/\bra{\hat n}_6\ket$ and
$\overline{Ke}=Ke/\bra{\hat n}_6\ket$, where $Coh$ and $Ke$ are
defined as
\begin{equation}\label{coh}
Coh=\bra {1\over{M(M-1)}}\sum_{i,j}^M {\hat a}_i^\dagger {\hat a}_j\ket
\end{equation}
\begin{equation}\label{ke}
Ke= \bra {1\over{2(M-1)}}\sum_i^{M-1} ({\hat a}_i^\dagger {\hat a}_{i+1}+{\hat a}_i {\hat a}_{i+1}^\dagger)\ket
\end{equation}
$Coh$ and $Ke$ show coherence between all lattice sites and between all the adjacent sites, respectively.
According to \esref{BHHsecond} and (\ref{coh}), between the expectation value of
the kinetic energy part of the Bose-Hubbard Hamiltonian
\begin{equation}\label{KE}
{KE} = - J \sum_i^{M-1} \bra({\hat a}_i^\dagger {\hat a}_{i+1}+{\hat a}_i {\hat a}_{i+1}^\dagger)\ket
\end{equation}
and $Ke$, \eref{ke}, which is an average of the coherence between adjacent cites,
there is a simple relation
\begin{equation}\label{keKE}
Ke={{-KE}\over{2J(M-1)}}, \quad M\neq 1
\end{equation}

As \fref{M11coh_ke_a_adags_tau_j_Rel} shows, the relative coherence
between sites 1 and 2, $\overline{\bra {\hat a}_1 {\hat
a}_{2}^\dagger\ket}$  , is different from that between sites 5 and
6, $\overline{\bra {\hat a}_5 {\hat a}_{6}^\dagger\ket}$,  due to
the edge (size) effect. Because the edge effect in a system with
size $M=11$ is small and all the other adjacent sites, except sites
$1$ and $2$, have almost the same coherence (or same relative
coherence) between each other as exists between sites $5$ and $6$,
we have, according to \fref{M11coh_ke_a_adags_tau_j_Rel}, at
temperatures $T_1=U$ and $T_2=0.5\hspace{0.1cm}U$ (within the
sampling error)
\begin{equation}\label{kea5a6dag}
 Ke=\bra {\hat a}_5 {\hat a}_{6}^\dagger\ket
\end{equation}
According to \eref{ke} and \fref{M11coh_ke_a_adags_tau_j_Rel}, for a
Bose-Hubbard model with M=11 sites in 1D, at temperatures $T_1=U$
and $T_2=0.5\hspace{0.1cm}U$, the kinetic energy of the system is
simply $KE=-20 J\bra {\hat a}_5 {\hat a}_{6}^\dagger\ket$.

In general, for $M$ lattice sites, the kinetic energy term $KE$, \eref{KE}, is given by
$-2 J(M-1)\bra {\hat a}_{M/2} {\hat a}_{(M+2)/2}^\dagger\ket$ for even $M$ and
$-2 J(M-1)\bra {\hat a}_{(M-1)/2} {\hat a}_{(M+1)/2}^\dagger\ket$ for odd $M\geq3$.

In \fref{M11coh_ke_a_adags_tau_j_Rel}, all the parameters show an
increase when either the system is cooled or the hopping matrix
element $J$ is increased.  By decreasing the temperature at a
constant $J$ the relative coherence increases; therefore the
ultracold bosons approach a superfluid state where the coherence
between the lattice sites is very high. When the constant
temperature is low enough, decreasing the tunnelling rate takes the
system to a Mott insulator phase where coherence between the sites
is lost~\cite{FisherM,Jaksch}.
\begin{figure*}%[p]
\begin{center}
$\begin{array}{ccc}
\includegraphics[angle=0,width=5.8cm,height=4.0cm]{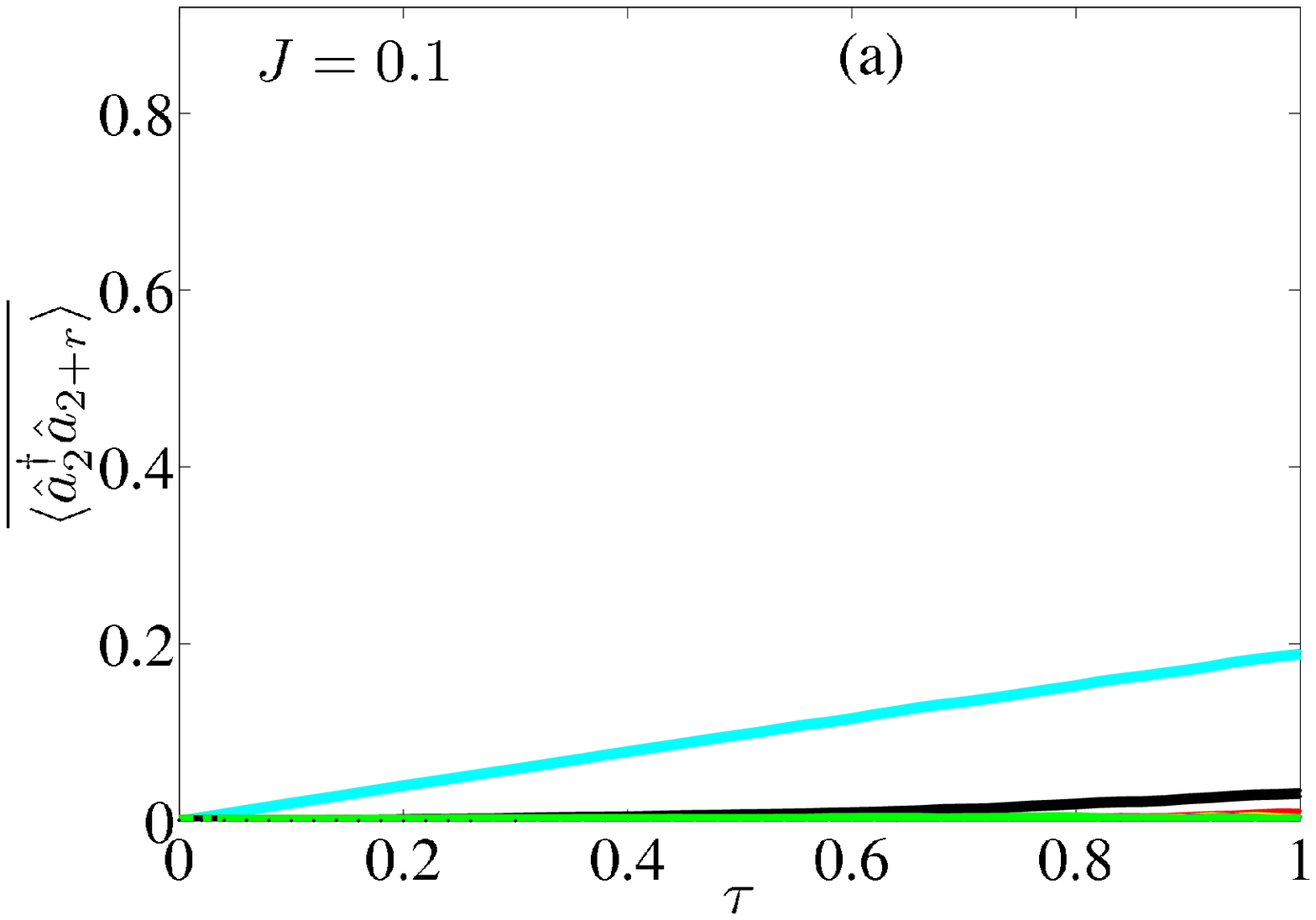}
\includegraphics[angle=0,width=5.8cm,height=4.0cm]{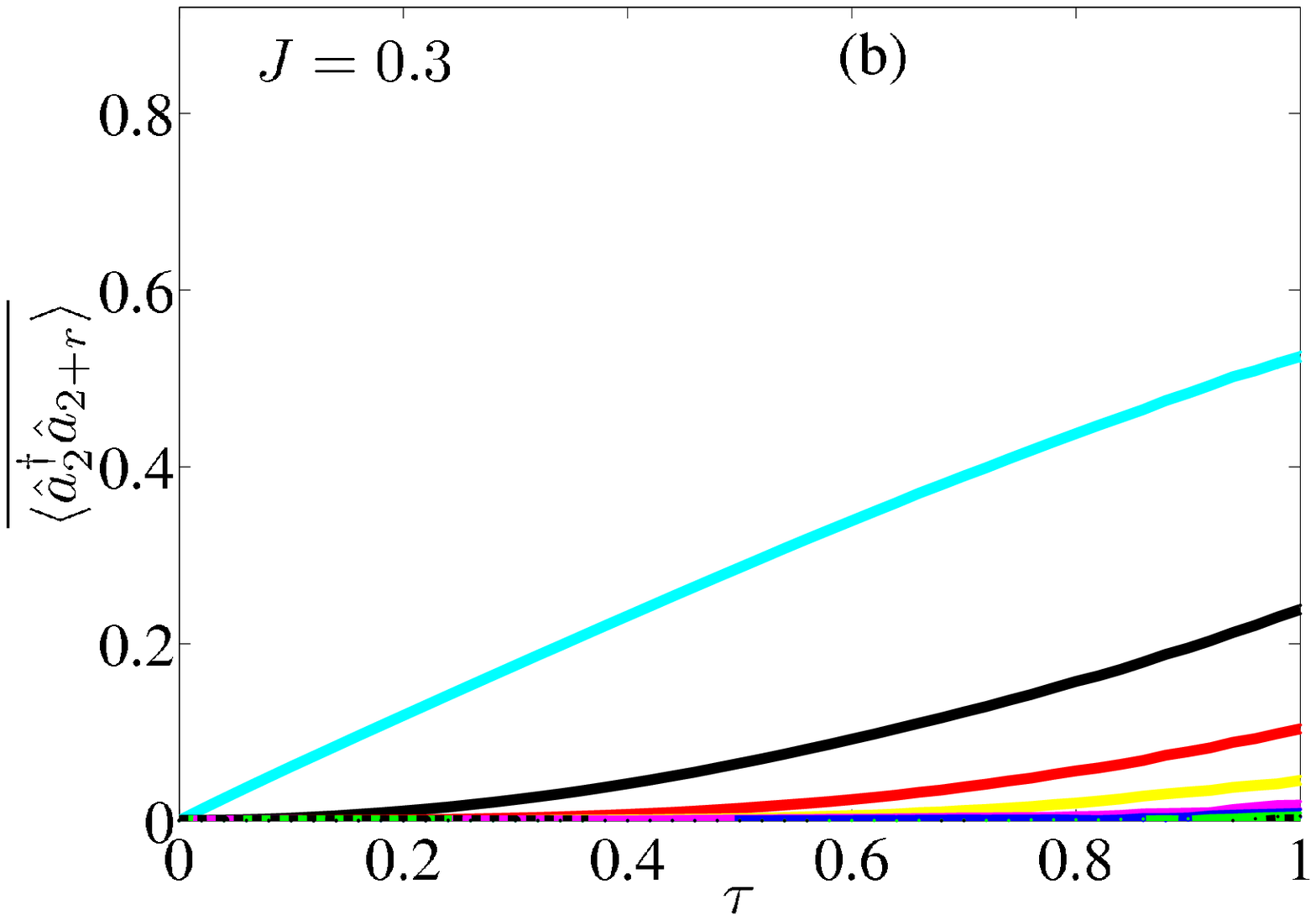}
\includegraphics[angle=0,width=5.8cm,height=4.0cm]{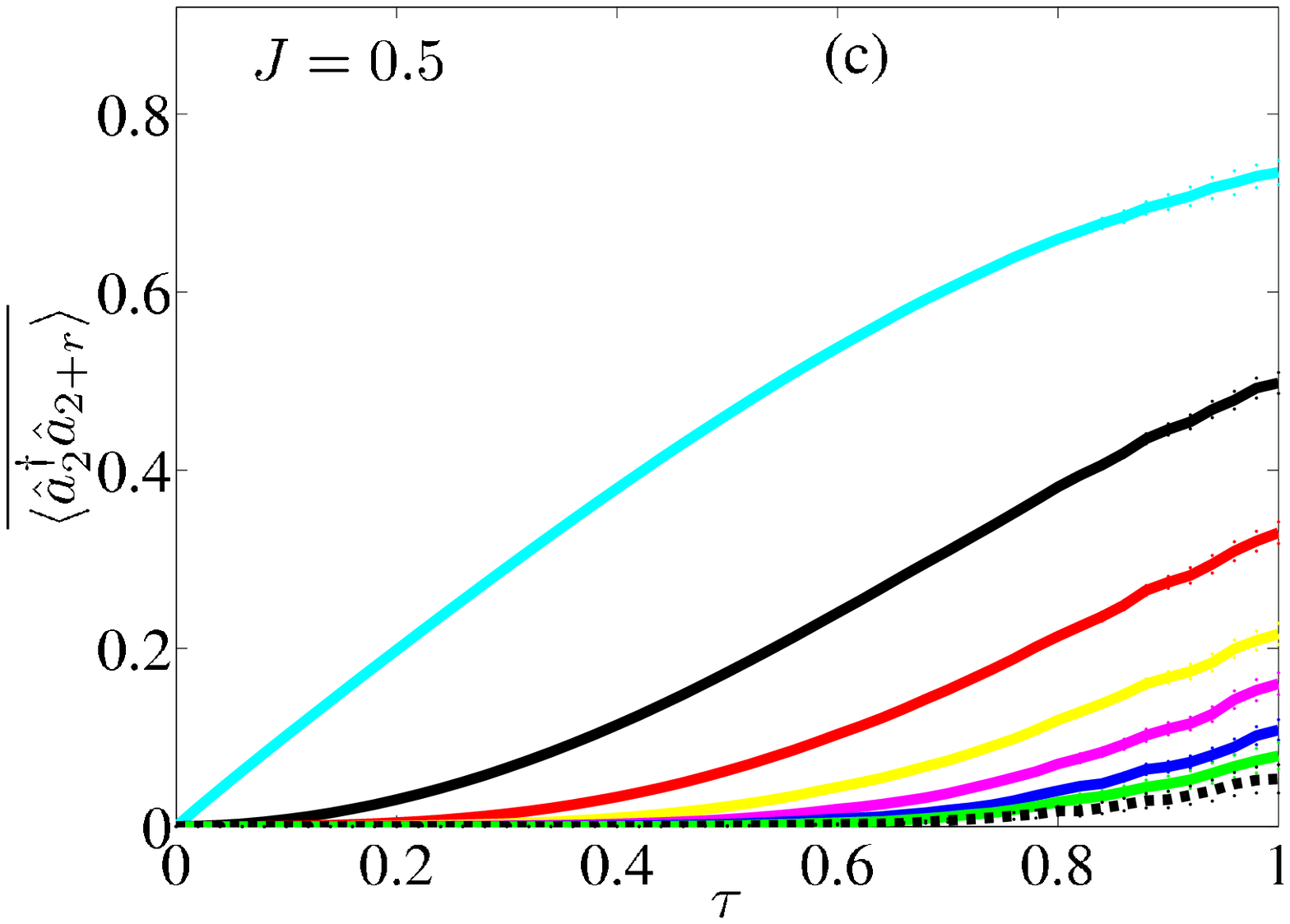}\\
\includegraphics[angle=0,width=5.8cm,height=4.0cm]{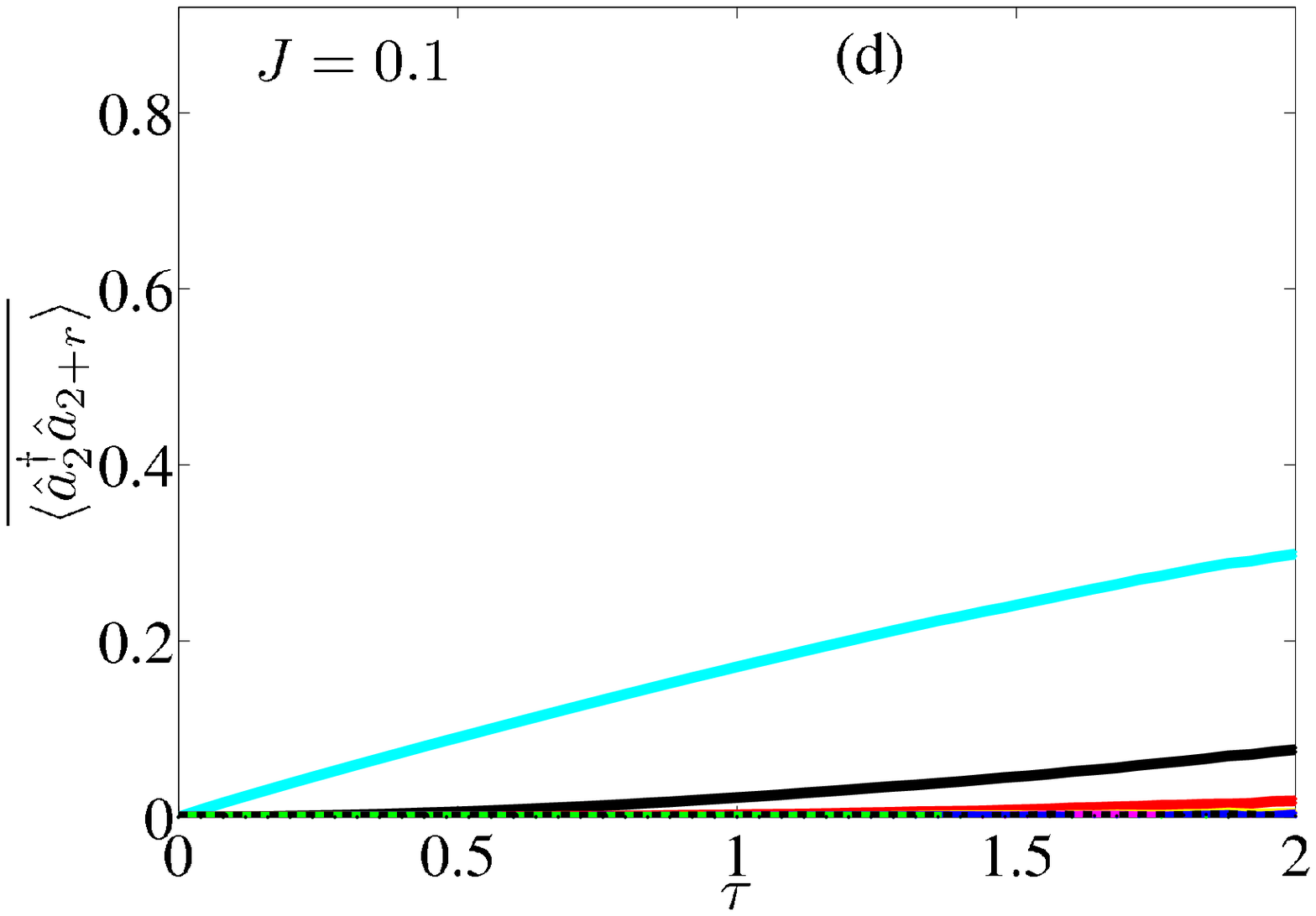}
\includegraphics[angle=0,width=5.8cm,height=4.0cm]{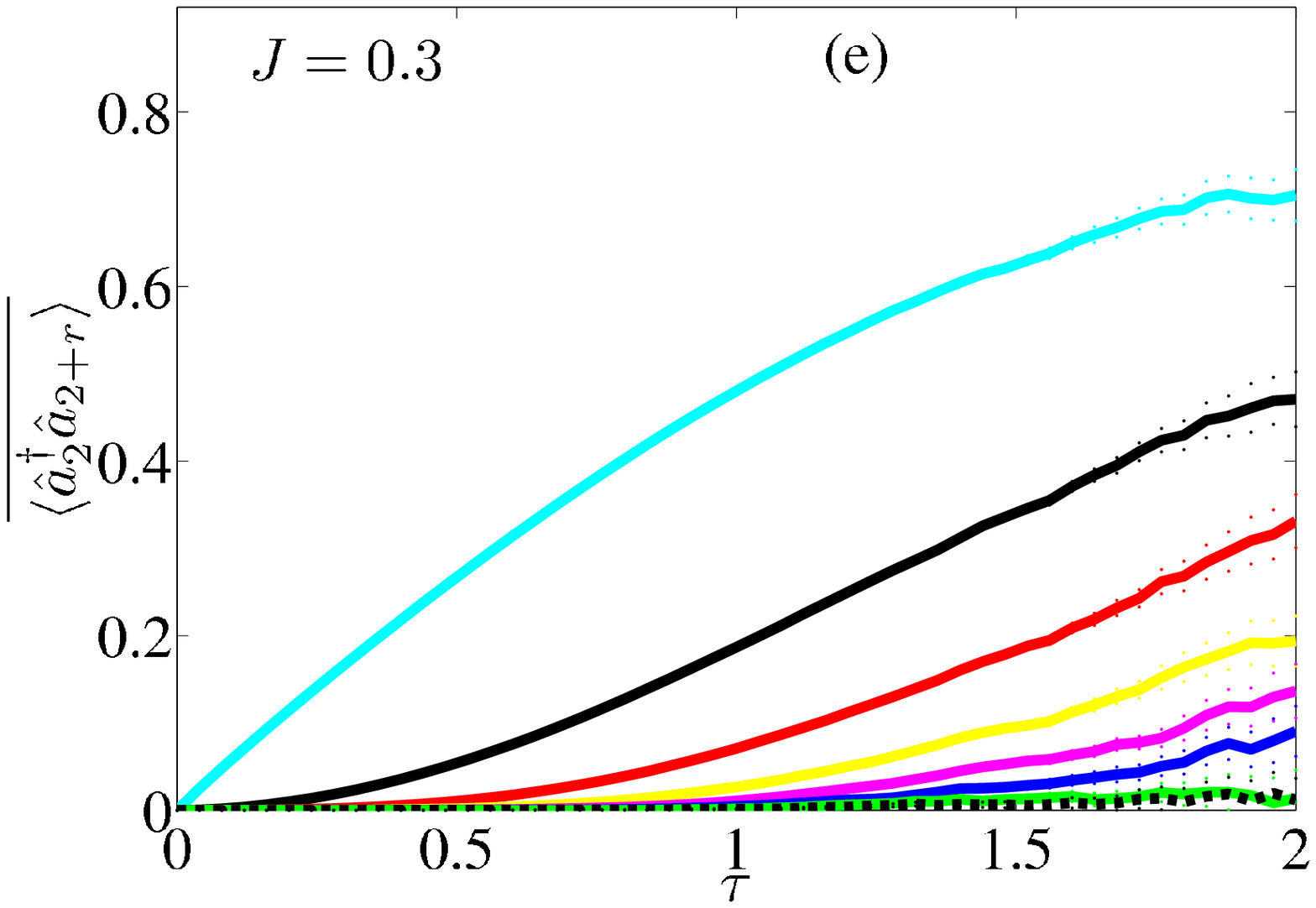}
\includegraphics[angle=0,width=5.8cm,height=4.0cm]{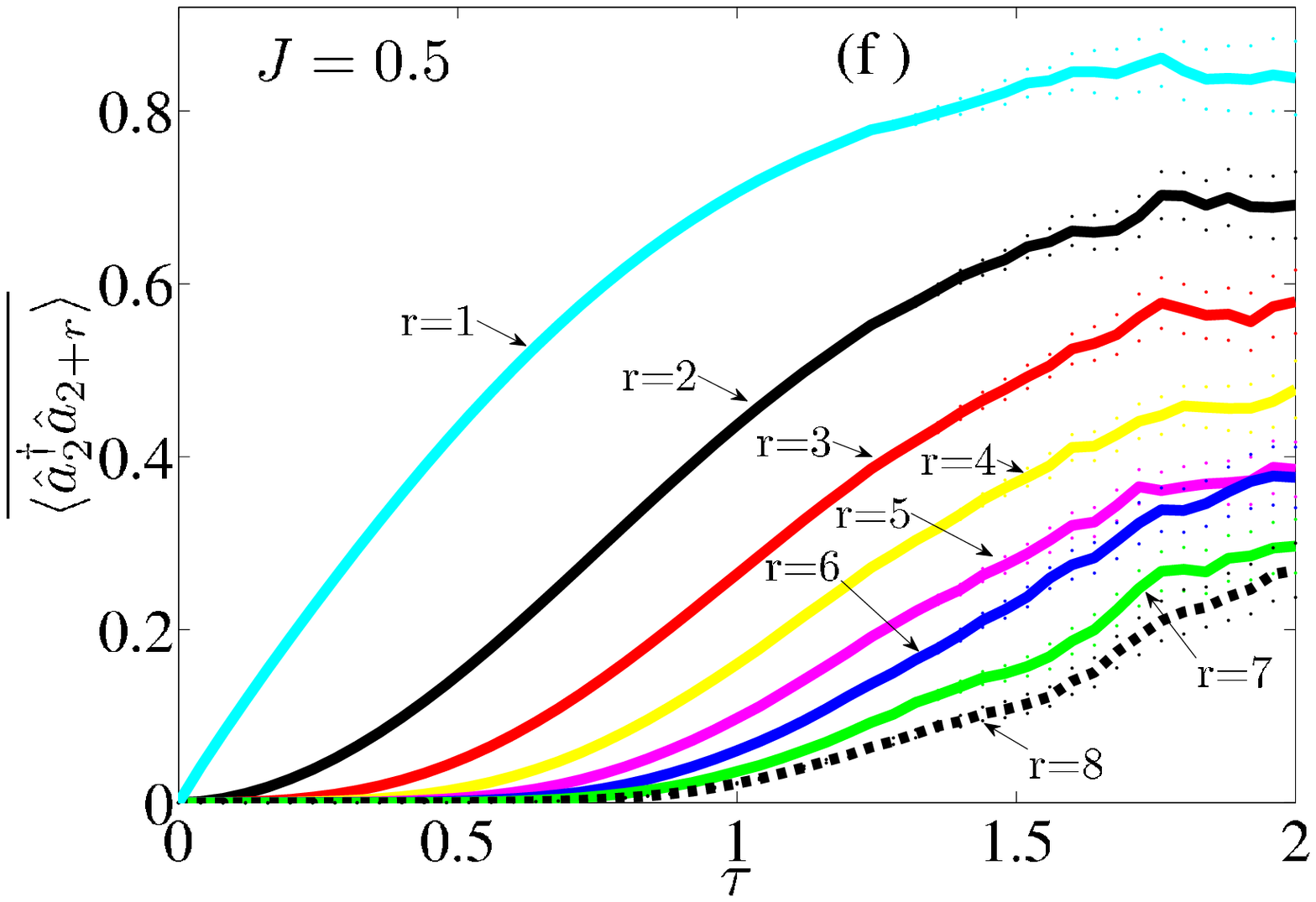}
\end{array}$
\caption[Relative atom-atom correlations $\overline{C_2(r)}=C_2(r)/\bra {\hat
n}_6\ket$ as a function of $\tau$]{\small Relative atom-atom
correlations $\overline{C_2(r)}=C_2(r)/\bra {\hat
n}_6\ket$ as a function of $\tau$
for M=11, $\mu_{{}_T}=0.5$, values of $J=0.3$ and $0.5$ and two
different target temperatures  $T=\hspace{0.1cm}U$ in (a) and (c)
and $T=0.5\hspace{0.1cm}U$ in (b) and (d).  The labels in (a)-(c)
are the same as in (d). The relative atom-atom correlations increase
when either the temperature is reduced or the hopping matrix element
is increased.}\label{a_dag_a_i_Rel}
\end{center}
\end{figure*}
\begin{figure*}%[th]
\begin{center}
$\begin{array}{cc}
\includegraphics[angle=0,width=8.5cm,height=5cm]{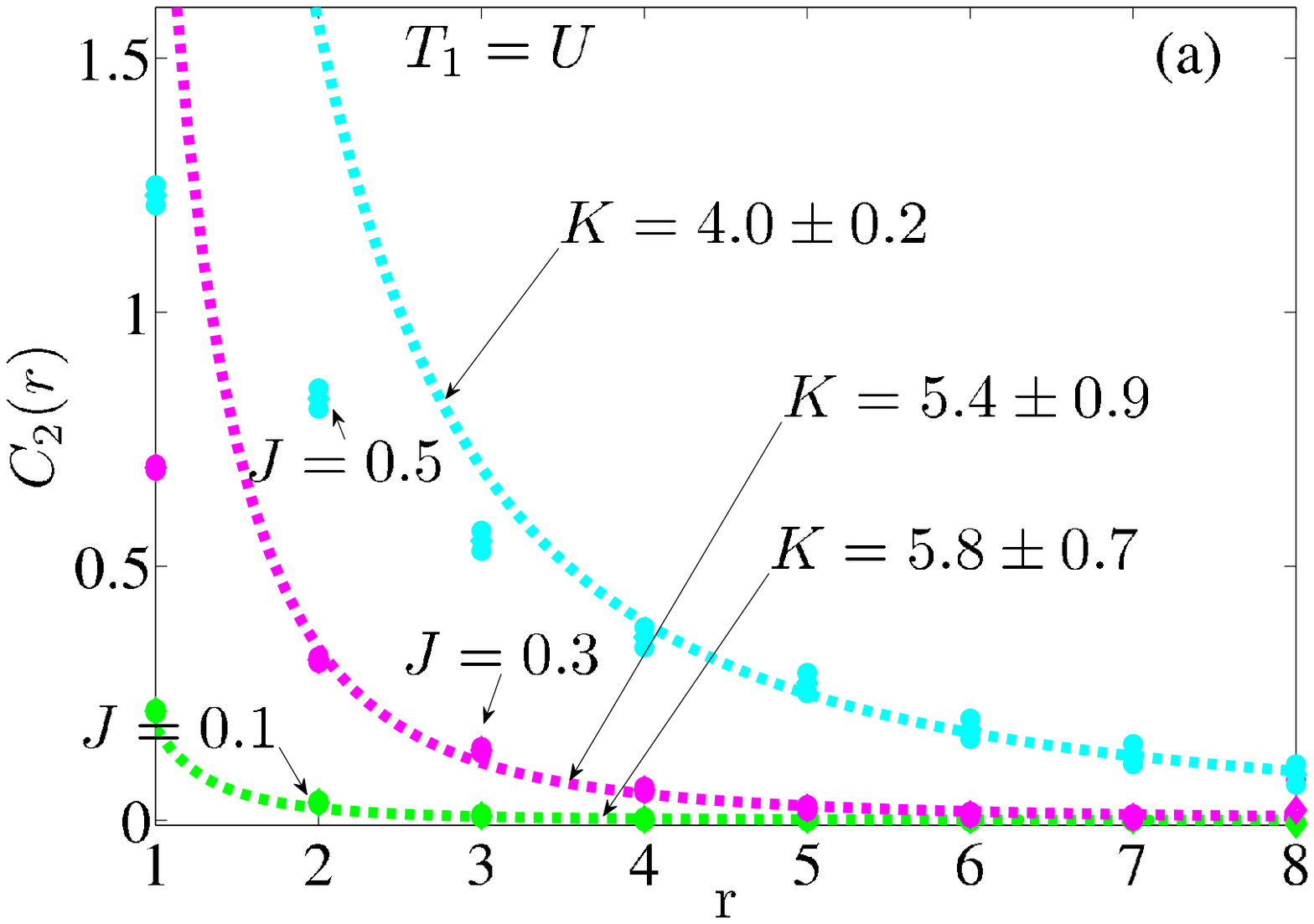}
\includegraphics[angle=0,width=8.5cm,height=5cm]{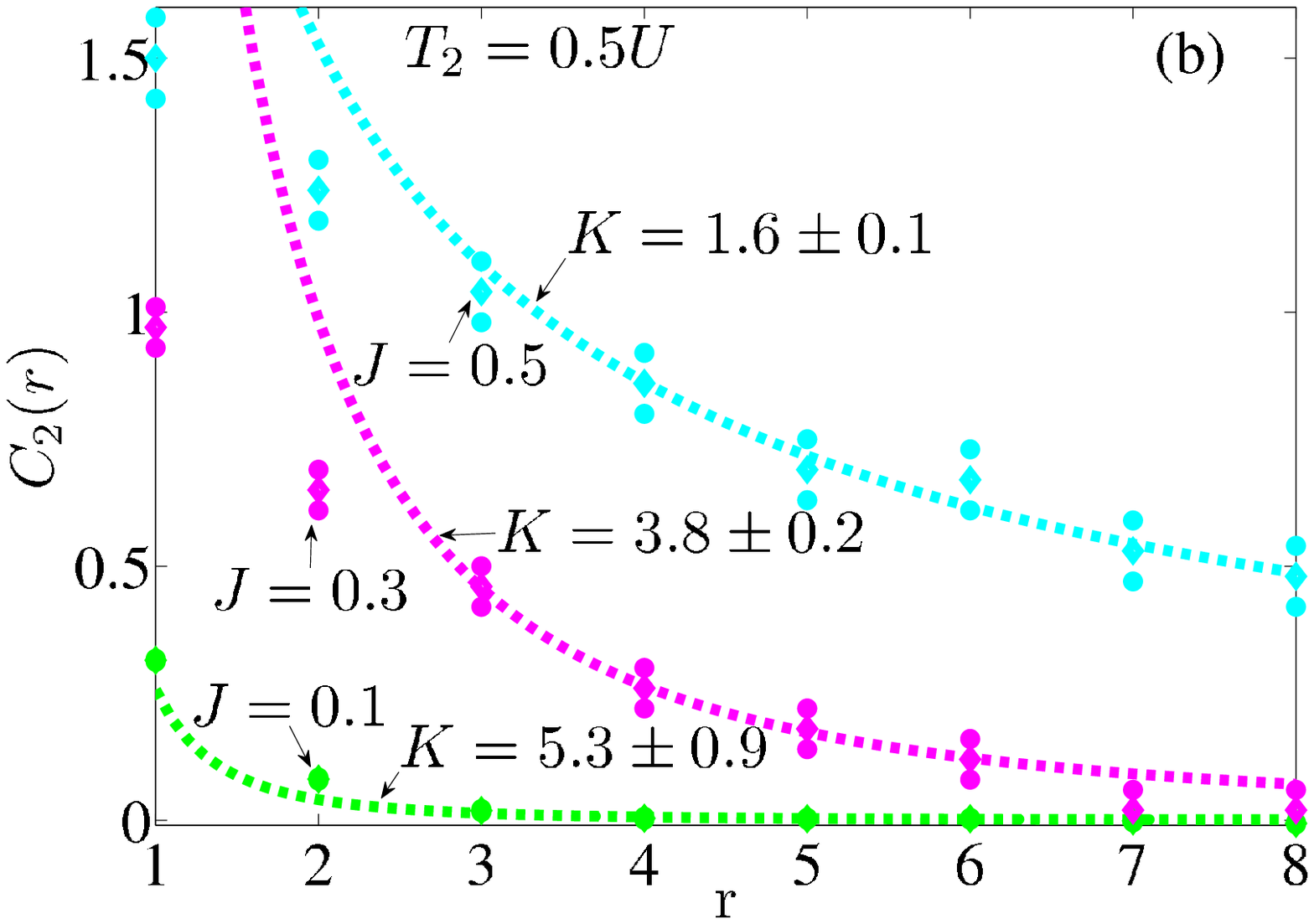}
\end{array}$
\caption[Atom-atom correlations $C_2(r)$ versus $r$]{\small
Atom-atom correlations $C_2(r)$ as a function of $r$, where $r$ is
the site number, for M=11, $\mu_{{}_T}=0.5$ and two different target
temperatures (a) $T=U$ and (b)
$T=0.5U$ and values of $J=0.1, 0.3$ and $0.5$.  Diamonds and
circles around them show the simulation results and their sampling
error, respectively. Dashed lines, which look like dotted lines, are
fits to the function $r^{-K/2}$ where $K$ is the Luttinger
parameter. According to this figure, $C_2(r)$ increases as
temperature is decreased. Also at a constant temperature by
increasing the hopping matrix element $J$, coherence between sites
increases.  Moreover, as $J$ is increased the Luttinger parameter
$K$ decreases.}\label{atomatom}
\end{center}
\end{figure*}
\Fref{coh_a5a6dag_n_delta_n} shows ${\bra{\hat n}_6 \ket}$ and the relative values $\overline{\Delta{n_6}}$, $\overline{\bra {\hat a}_5 {\hat
a}_6^\dag\ket}$, $\overline{Coh}$, $J\overline{\bra{\hat n}_6 \ket}$, $J\overline{\Delta{n_6}}$,
$J\overline{\bra {\hat a}_5 {\hat a}_6^\dag\ket}$ and  $J \overline{Coh}$ as a function of $J$ for two different
temperatures.  Here again, according to the figure, all the
parameters increase when either the temperature is reduced or the
hopping matrix element $J$ is increased.

\Fref{a_dag_a_i_Rel} shows $\overline{C_2(r)}=C_2(r)/\bra {\hat
n}_6\ket$ as a function of $\tau$. All values of $\overline{C_2(r)}$
show an increase when $T$ is reduced. The same behaviour is observed
when $J$ is increased.

According to \fsref{M11coh_ke_a_adags_tau_j_Rel} and
\ref{a_dag_a_i_Rel}, all relative measures of the coherence $\overline{Coh}$, $\overline{Ke}$,
$\overline{\bra {\hat a}_1 {\hat a}_{2}^\dagger\ket}$,
$\overline{\bra {\hat a}_5 {\hat a}_{6}^\dagger\ket}$ and
$\overline{C_2(r)}$ increase when either the temperature is
reduced or the hopping matrix element is increased.
\section{Calculation of the Luttinger parameter}
The important parameter which has been used to locate the critical
parameters of the superfluid to Mott insulator quantum phase
transition  is atom-atom correlations $C(r)=\bra{\hat a}^{\dagger}_0
{\hat a}^{}_{r}\ket$~\cite{Kuhner98,Kuhner00} and more generally
$C_i(r)=\bra{\hat a}^{\dagger}_i  {\hat a}^{}_{i+r}\ket$ against
$J$~\cite{Damski06}.

According to Refs.~\cite{GiamarchiMillis92,Giamarchi97,Glazman97}, basically,
it is possible to fit the function  $r^{-K/2}$ to the curve
of  $C_2(r)$ as a function of $r$ for different values of the chemical potential and hopping matrix element,
 find the boundaries of the Mott lobes, and obtain the phase diagram of the Bose-Hubbard model.
 \Fref{atomatom} shows the atom-atom correlations $C_2(r)$ as a
function of $r$ for $T=U$ and $T=0.5U$. At each
temperature three  different values of the hopping matrix element
$J$ are considered. According to this figure, $C_2(r)$ increases as
the temperature is decreased. Also at a constant temperature by
increasing the hopping matrix element $J$, the coherence between
sites increases.  Moreover, as $J$ is increased the Luttinger
parameter $K$ decreases.
At $\mu_{{}_T}/U=0.5$ and at the constant temperatures $T_1=U$ and
$T_2=0.5\hspace{0.1cm}U$, by increasing the hopping matrix element
$J$, relative atom-atom correlations $\overline{C_2(r)}$, where $C_2(r)=\bra{\hat a}^{\dagger}_2  {\hat
a}^{}_{2+r}\ket$, which are other measures of the relative coherence between
sites, increase but the Luttinger parameter $K$, which is important
in locating the boundaries of the Mott-insulator lobes, decreases.
Also, for a constant value of the hopping matrix element $J=0.5$, by
reducing the temperature from $T_1=U$ to $T_2=0.5\hspace{0.1cm}U$,
the Luttinger parameter $K$ changes from $4.0\pm 0.2$ to $1.6\pm
0.1$.
\section{Conclusion}
In this paper, using the Bose-Hubbard model, we simulated ultracold
atoms  in the grand canonical ensemble of quantum degenerate gases.
We studied ultracold atoms at finite temperatures with the gauge $P$
representation. We have written a simulation code using XMDS, which generates a C++ code.  In 1D we simulated
the Bose-Hubbard model with 1, 2, 3, 7 and 11 sites.

The simulation results are in good agreement with highly
accurate numerical calculations, based on a truncated number-state
basis, when there is no on-site interaction between atoms. Also, for
a double well system, we compared the simulation results with
exact  analytical results for the case where the atoms can tunnel
between sites but the on-site interaction between the sites is zero.

We also investigated the average number of particles, relative standard
deviations and coherences between sites at finite temperatures for
the Bose-Hubbard model in 1D with open boundary conditions
consisting of 1, 2, 3, 7 and 11 sites  and showed that at non-zero
temperatures the relative standard deviation is not zero even for $J/U=0$ and
grows as $J/U$ is increased.  We found that the relative standard deviation
$\overline{\Delta n_i}$ is higher at higher temperatures and is less than the
corresponding relative standard deviation for a coherent state and is much
less than that of a thermal state with the same value of $\bra {\hat
n}_i \ket$.

For $J=0$, we showed that above $T=0.1 \hspace{0.1cm}U$ the stepwise
pattern in the plot of $\bra{\hat n}_i \ket$ versus the target
chemical potential $\mu_{{}_T}$ starts to vanish; therefore, there
is no Mott insulator-like lobe in the phase diagram of the
Bose-Hubbard model in the plane of $\mu/U$-$J/U$.

Comparing the Bose-Hubbard model with $M=3$, $7$ and $11$,  we
showed that in a 1D Bose-Hubbard model with $M=11$ and open boundary
conditions edge effects are insignificant except for the side sites
(the first and last sites).

At low temperatures, for constant values of $J/U=0.2$ and
$\mu_{{}_T}/U=0.5$, by reducing the temperature from $T_1=U$ to
$T_2=0.5\hspace{0.1cm}U$, the relative standard deviation of the number of
particles at each lattice $\overline{\Delta n_i}$ decreases but remains well
below those of a coherent state, $\overline{\Delta n_{coh}}$, and also a
thermal state, $\overline{\Delta n_{th}}$, with the same average number of
particles. The relative standard deviation $\overline{\Delta n_i}$ is closer to  $\overline{\Delta
n_{coh}}$ than to $\overline{\Delta n_{th}}$.  This confirms that the ground
state of a superfluid can be described by a coherent state.

For $\mu_{{}_T}/U=0.5$, at the constant temperatures $T_1=U$ and
$T_2=0.5\hspace{0.1cm}U$, the relative standard deviation $\overline{\Delta n_i}$
decreases as $J/U$ is reduced.  At these temperatures the lowest
value of the relative standard deviation  of the number of particles at each
lattice $\overline{\Delta n_i}$ is not zero even if the tunnelling is zero.
This confirms that at these temperatures,  for $J/U=0$,  the
compressibility $\kappa$ is not zero, so there is no Mott insulator
phase present at these temperatures.  This is in good agreement with
the temperature $T=0.1 \hspace{0.1cm}U$ discussed above or, more
accurately, with the melting temperature $T_0=0.06 \hspace{0.1cm}U$
 above which there is no Mott insulator phase in the Bose-Hubbard model~\cite{GhanbariThesis}.

At the constant temperatures $T_1=U$ and $T_2=0.5\hspace{0.1cm}U$,
for $\mu_{{}_T}/U=0.5$, by increasing the tunnelling rate all
physical quantities $\overline{Coh}$, $\overline{Ke}$, $\overline{\bra {\hat a}_1 {\hat
a}_{2}^\dagger\ket}$, $\overline{\bra {\hat a}_5 {\hat a}_{6}^\dagger\ket}$ and
$\overline{C_2(r)}$, which measure $relative$ $coherence$ between the lattice sites,
increase.  Also, if $J/U$ is kept constant but the temperature is
reduced from $T_1=U$ to $T_2=0.5\hspace{0.1cm}U$, all of them
increase.  Likewise, all measures of $coherence$
 increase when either the temperature is reduced or the
hopping matrix element is increased.

For a Bose-Hubbard model with M=11 sites in 1D, at temperatures
$T_1=U$ and $T_2=0.5\hspace{0.1cm}U$, the kinetic energy part of the
system is simply given by $-20 J\bra {\hat a}_5 {\hat
a}_{6}^\dagger\ket$, within the sampling error, which can be
generalised to  $KE=-2 J(M-1)\bra {\hat a}_{M/2} {\hat
a}_{(M+2)/2}^\dagger\ket$ ,for an even number of lattice sites M,
and to $KE=-2 J(M-1)\bra {\hat a}_{(M-1)/2} {\hat
a}_{(M+1)/2}^\dagger\ket$ for an odd number of lattice sites M.

At $\mu_{{}_T}/U=0.5$ and at the constant temperatures $T_1=U$ and
$T_2=0.5\hspace{0.1cm}U$, by increasing the hopping matrix element
$J$, the Luttinger parameter $K$, which is important
in locating the boundaries of the Mott-insulator lobes, decreases.
Also, by
reducing the temperature from $T_1=U$ to $T_2=0.5\hspace{0.1cm}U$, for a constant value of the hopping matrix element $J=0.5$ the Luttinger parameter $K$ changes from $4.0\pm 0.2$ to $1.6\pm
0.1$.
%Provided new gauges are proposed to decrease the sampling error at low temperatures,
% the techniques described in this paper could be extended to
%investigate a number of other interesting topics, including:

With the particular gauge-choice used here, we have found that the the growth of
sampling error is a limiting factor at low temperatures and a large number of sites.
  However, it is always possible that for particular situations a better choice of
gauge may reduce the sampling error.

Even with the sampling error limitations described above, the gauge-P method is very
general and could potentially be applied to a range of ultracold lattice systems,
including

\begin{enumerate}
\item The $J/U$ critical ratios and phase diagrams in 1D, 2D, and 3D,
for the Bose-Hubbard model and also the critical
values and phase diagrams of the superfluid to Mott insulator quantum
phase transition at zero temperature.
\item Disordered Bose-Hubbard model~\cite{FisherM,Krutitsky06,Roth03,buonsante:011602}
 and also two component bosons in periodic lattices~\cite{Roth03,Altman03}.
\item The coexistence of the Mott insulator and superfluid phases in inhomogeneous traps,
such as quadratic and quartic trapping potentials, for a continuous range
 of incommensurate fillings~\cite{BatrouniRousseau02,Gygi__Troyer}.
\item Ultracold bosons in a double-well potential~\cite{Milburn97} and a
tilted multi-level double-well potential~\cite{DounasarXiv07}.
\item Extended Bose-Hubbard models~\cite{pai:104508}.
\item Strongly interacting bosons in a 2D rotating square lattice which can
also be studied via a modified Bose-Hubbard Hamiltonian~\cite{Bhat06}.
\end{enumerate}
\appendix
\section{Positive $P$ representation}\label{pp_rep}
There are density operators for which the  $P$ representation  does
not exist. For example, the $P$ representation
cannot describe nonclassical effects such as squeezing or
antibunching~\cite{GardinerZoller}. In contrast, the positive $P$ representation gives
stochastic differential equations which can represent genuine
quantum-mechanical problems like squeezing or
antibunching~\cite{CarterReid,DrummondShelby,DrummondFicek}.

In  the positive $P$ representation, we can
write~\cite{DrummondGardiner}
\begin{equation}\label{ro}
{{\hat\rho}}= \int P({\bm{\alpha}},{\bm{\beta}},\tau) {\hat{\Lambda} } d^{4M}
\vec{\lambda}
\end{equation}
where $\vec{\lambda}=({\bm{\alpha}},{\bm{\beta}})=({\alpha_1, \alpha_2, \cdots,
\alpha_M},{\beta_1, \beta_2, \cdots,
\beta_M})$ and
\begin{equation}
{\hat\Lambda} ={ {{|{{{\bm{\alpha}}}}\rangle\langle{{{\bm{\beta}}}}^\ast
|}}\over{\langle{{{{{\bm{\beta}}}^\ast}}}|{{{\bm{\alpha}}}} \rangle}}=|{ {{|{{{\bm{\alpha}}}}\rangle\langle{{{\bm{\beta}}}}^\ast
|}}}|e^{-{\bm{\alpha}}\cdot{\bm{\beta}}}
\end{equation}
$||\bm{\alpha}\rangle$ is an M-dimensional Bargmann
coherent state~\cite{GardinerZoller} and
$||\alpha_i\rangle =e^{|\alpha_i|/2}|\alpha_i\rangle$,
where $M$ is the number of modes (number of lattice sites, for
example), $\tau$ may represent real time or imaginary time (inverse
temperature) and
\begin{equation}
|\alpha_i\rangle = exp \left( -{1\over 2} |\alpha_i|^2\right) \sum_{n=0}^\infty {{\alpha_i^n}\over{n!}}|n\rangle,
\end{equation}
where $i=1, 2, \cdots, M$. We also have the identities
%{\hat\Lambda} = \Omega {\partial \over {\partial
%\Omega}} {\hat\Lambda}, \quad
\begin{eqnarray}\label{identities}
 \hat{\bf a} {\hat\Lambda} &=& {{\bm{\alpha}}} {\hat\Lambda}, \quad {\hat{\bf a}}^\dag {\hat\Lambda} =
\left[{{\bm{\beta}}} + {\partial \over {\partial {{\bm{\alpha}}}}}\right]
{\hat\Lambda}, \nonumber
\\  {\hat\Lambda} {\hat{\bf a}} &=& \left[{{\bm{\alpha}}} + {\partial \over {\partial {{\bm{\beta}}}}}\right]
{\hat\Lambda}, \quad  {\hat\Lambda} {\hat{\bf a}}^\dag = {{\bm{\beta}}}
{\hat\Lambda}.
\end{eqnarray}
Expectation values of the normally ordered products ${\hat
a}_i^{\dag m}{\hat a}_i^n$ can be written in the positive $P$
representation as the following
\begin{equation}
\bra {\hat a}_i^{\dag m} {\hat a}_i^n \ket = \int {\beta}_i^m {\alpha}_i^n P({\bm{\alpha}},{\bm{\beta}},\tau) {\hat{\Lambda} } d^{4M}
\vec{\lambda}
\end{equation}
Also, after integration by parts, provided the boundary terms vanish at infinity, we can write
\begin{equation}
\int {{\partial P({\bm{\alpha}},{\bm{\beta}},\tau)}\over{\partial\tau}}
 {\hat{\Lambda} } d^{4M} \vec{\lambda} =   \int   P({\bm{\alpha}},{\bm{\beta}},\tau)
{{{{\cal L}}^{(+)}_A}} {\hat{\Lambda} } d^{4M} \vec{\lambda}
 \end{equation}
When there are no terms higher than second order, ${{{ {\cal L}}^{(+)}_A}}$
may be expanded as
\begin{equation}\label{AVD0}
{{{{\cal L}}^{(+)}_A}} = V  + A^{(+)}_j \partial_j + {1\over 2} D_{ij}
\partial_i\partial_j
\end{equation}
In the positive $P$ representation, the Fokker-Planck equation is
given by
\begin{equation}\label{FP}
{{\partial P({\bm{\alpha}},{\bm{\beta}},\tau)}\over{\partial\tau}} =
(V - \partial_j A^{(+)}_j  + {1\over 2}\partial_i\partial_j D_{ij})
P({\bm{\alpha}},{\bm{\beta}},\tau)
\end{equation}
\section{Gauge $P$ representation}\label{gp_rep}
In  the gauge $P$ representation, the density matrix can be written
as~\cite{DeuarDrummond}
\begin{equation}\label{rogauge}
{{\hat\rho}}= \int G(\vec{\alpha},\tau) {\hat{\Lambda} } d^{4M+2}
\vec{\alpha}
\end{equation}
where $\vec{\alpha}=(\alpha^0, \alpha^1, \cdots, \alpha^M, \alpha^{M+1}, \alpha^{M+2}, \cdots, \alpha^{2M})=(\Omega, {\bm{\alpha}}, {\bm{\beta}})$, and
\begin{equation}
{\hat\Lambda} =\Omega{ {{|{{{\bm{\alpha}}}}\rangle\langle{{{\bm{\beta}}}}^\ast
|}}\over{\langle{{{{{\bm{\beta}}}^\ast}}}|{{{\bm{\alpha}}}} \rangle}}=\Omega|{ {{|{{{\bm{\alpha}}}}\rangle\langle{{{\bm{\beta}}}}^\ast
|}}}|e^{-{\bm{\alpha}}\cdot{\bm{\beta}}}
\end{equation}
When $\Omega=1$, this phase space representation reduces to the positive $P$ representation~\cite{DeuarDrummond}.
Here, we have the identities given by~\eref{identities} plus $\Omega {\partial \over {\partial\Omega}}{\hat\Lambda}={\hat\Lambda}$.

In this phase space representation, quantum averages of the normally ordered products ${\hat a}_i^{\dag m}{\hat a}_i^n$ are
\begin{equation}
\bra {\hat a}_i^{\dag m} {\hat a}_i^n \ket =  {{\bra{\beta}_i^m
{\alpha}_i^n \Omega + ({\alpha}_i^m {\beta}_i^n
\Omega)^\ast\ket_{stoch}}\over{\bra\Omega +\Omega^\ast\ket_{stoch}}}
\end{equation}
where
$
{\bra f(\bm{\alpha})\ket_{stoch}}=\int f(\bm{\alpha}) G(\vec{\alpha},\tau)  d^{4M+2}
\vec{\alpha}
$.

After integration by parts, provided the boundary terms vanish at infinity, we have
\begin{equation}
\int {{\partial G(\vec{\alpha},\tau)}\over{\partial\tau}}
 {\hat{\Lambda} } d^{4M} \vec{\alpha} =   \int   G(\vec{\alpha},\tau)
{{{{\cal L}}_{{}_{GA}}}} {\hat{\Lambda} } d^{4M+2} \vec{\alpha}
 \end{equation}
In the gauge $P$ representation, when there are no terms higher than
second order, ${\cal L}_{{}_{GA}}$ may be expanded as
\begin{equation}\label{LGA}
{{{{\cal L}}_{{}_{GA}}}}={{{{\cal L}}^{(+)}_A}} + \left[ V+ \frac{1}{2} {\vec g}\cdot{\vec g}
\Omega \partial_\Omega + g_k B_{jk} \partial_j \right]  (\Omega \partial_\Omega -1)
\end{equation}
\begin{equation}\label{LGA1}
{{{{\cal L}}_{{}_{GA}}}}=A_\mu \partial_\mu + \frac{1}{2}
{\underline{\underline{D}}}_{\mu\nu} \partial_\mu \partial_\nu,
\quad \mu, \nu=0, 1, 2, \cdots, 2M
\end{equation}
where ${{{{\cal L}}^{(+)}_A}}$ is given by \eref{AVD0}, ${\vec g} = \{g_i(\vec{\alpha})\}$ are 2M arbitrary gauge functions and
\begin{equation}\label{vecA}
{\vec A}=(A_0, A_1, \cdots, A_{2M}), \quad A_0=\Omega V, \quad A_j=A^{(+)}_j-g_k B_{jk}
\end{equation}
\begin{equation}\label{D_BBT}
{\bf D}={\bf B}{\bf B}^T, \quad \underline{\underline{\bf D}}=\underline{\underline{\bf B}} \hspace{0.1cm} \underline{\underline{\bf B}}^T, \quad \underline{\underline{\bf B}} =  \left( \begin{array}{cc}
0 & \Omega {\vec g}  \\
0 & {\bf B}  \\
\end{array} \right)
\end{equation}
Now, the \^{I}to form~\cite{Gardiner} of the Langevin  equations are
\begin{equation}\label{ItoOmega}
d\Omega = \Omega \left(V d\tau + \sum_{k=1}^{2M} g_k dW_k \right),
\end{equation}
\begin{equation}\label{ItoAlpha}
d\alpha^j = \left(A^{(+)}_j-\sum_{k=1}^{2M} g_kB_{jk}\right) d\tau +\sum_{k=1}^{2M} B_{jk} dW_k,
\end{equation}
where $j=1, 2, \cdots, 2M$ and  the Wiener increments $dW_i$ have
the property $\bra dW_i(\tau)dW_j(s)\ket_s
=\delta_{ij}\delta(\tau-s)d\tau^2$~\cite{DrummondGardiner,DeuarDrummondI}
which can be realized at each $d \tau$ by real Gaussian noises with
zero average and variance $d \tau$.
\section{\^{I}to and Stratonovich forms of the Langevin equations}\label{ItoStra}
Considering \esref{ItoOmega} and (\ref{ItoAlpha}) for the Bose-Hubbard model, \^{I}to Langevin equations are
%\begin{widetext}
\begin{eqnarray}\label{Omega}
d\Omega &=& \Omega \left(J\sum_{i,j}^M \omega_{ij}\alpha_i {\bf{\beta}}_j  -\frac{U}{2}
\sum_{i=1}^M n_i^2  +\mu_e \sum_{i=1}^M
n_i \right) d\tau \nonumber \\ &+& \Omega \sum_{k=1}^{2M} g_k dW_k
\end{eqnarray}
\begin{eqnarray}
d\alpha^j &=& \left[\frac{J}{2} \sum_{i=1}^{2M} {\omega_{ji}} \alpha^i -
\frac{U}{2}(|n_j|+in^{\prime\prime}_j)\alpha^j +
\frac{\mu_e}{2}\alpha^j\right] d\tau  \nonumber \\ &+& i \sqrt{\frac{U}{2}}\alpha^j
dW_j,  \quad j=1, 2, \cdots, 2M.
\end{eqnarray}
%\end{widetext}
The Stratonovich stochastic equations are~\cite{Gardiner}
\begin{equation}\label{StAlpha}
d\alpha^{\mu (S)} = \left(A^{(S)}_\mu\right) d\tau +\sum_{k=1}^{2M} B_{\mu k} dW_k,
\end{equation}
where $ A^{(S)}_\mu= A_\mu -\frac{S_\mu}{2},  \quad \mu=0, 1, 2, \cdots, 2M$ and
\begin{equation}\label{Smu}
S_\mu= \sum_{\nu=0}^{2M}  \sum_{\gamma=0}^{2M} \left( \underline{\underline{\rm B}}_{\gamma\nu}\partial_{\alpha^\gamma} +
{\underline{\underline{\rm B}}^*_{\gamma\nu}}\partial_{{\alpha^{\gamma}}^*} \right)\underline{\underline{\rm B}}_{\mu\nu}
\end{equation}
Because $\underline{\underline{\rm B}}_{\mu0}$ is zero, we can write
\begin{equation}\label{Si}
S_\mu= S^1_\mu+S^2_\mu
\end{equation}
where
\begin{equation}\label{Si1}
S^1_\mu= \sum_{j=1}^{2M} \left( \underline{\underline{\rm B}}_{0 j}\partial_{\Omega} +
{\underline{\underline{\rm B}}^*_{0 j}}\partial_{\Omega^*} \right)\underline{\underline{\rm B}}_{\mu j}
\end{equation}
\vspace{-0.5cm}
\begin{equation}\label{Si2}
S^2_\mu= \sum_{j=1}^{2M}  \sum_{k=1}^{2M} \left( \underline{\underline{\rm B}}_{k j}\partial_{\alpha^k} +
{\underline{\underline{\rm B}}^*_{kj}}\partial_{{\alpha^{k}}^*} \right)\underline{\underline{\rm B}}_{\mu j}
\end{equation}
We have $\underline{\underline{\rm B}}_{0j}=\Omega g_j, \hspace{0.2cm}\partial_{\Omega}
\underline{\underline{\rm B}}_{0 j}=g_j, \hspace{0.2cm}\partial_{\Omega^*} \underline{\underline{\rm B}}_{0 j}=0,
\hspace{0.2cm}\underline{\underline{\rm B}}_{i j}=i\sqrt{\frac{U}{2}}\delta_{ij}\alpha^j$, so
\begin{equation}\label{Si1_4}
S^1_0= \Omega\sum_{j=1}^{2M} g^2_j
\end{equation}
Also
%\begin{widetext}
\begin{equation}\label{Si2_1}
S^2_0=\sum_{j=1}^{2M}  \sum_{k=1}^{2M} \left( \underline{\underline{\rm B}}_{k j}\partial_{\alpha^k} +
{\underline{\underline{\rm B}}^*_{kj}}\partial_{{\alpha^{k}}^*} \right)\underline{\underline{\rm B}}_{0 j} \nonumber
\end{equation}
\vspace{-0.5cm}
\begin{equation}
=i\Omega\sqrt{\frac{U}{2}} \sum_{j=1}^{2M}  \sum_{k=1}^{2M} \left(
\delta_{kj}\alpha^j\partial_{\alpha^k}   - \delta_{kj}{\alpha^j}^*\partial_{{\alpha^k}^*}  \right) g_j   \nonumber
\end{equation}
\vspace{-0.5cm}
\begin{equation}
= i\Omega\sqrt{\frac{U}{2}}\sum_{k=1}^{2M} \left(
\alpha^k\partial_{\alpha^k}   - {\alpha^k}^*\partial_{{\alpha^k}^*}  \right) g_k \nonumber
\end{equation}
\vspace{-0.5cm}
\begin{equation}
= i\Omega\sqrt{\frac{U}{2}}\sum_{k=1}^{M}
\left[ (\alpha_k\partial_{\alpha_k}+\beta_k\partial_{\beta_k})
-({\alpha_k}^*\partial_{{\alpha_k}^*}+{\beta_k}^*\partial_{{\beta_k}^*})   \right] g_k \nonumber
\end{equation}
\vspace{-0.5cm}
\begin{equation}
= i\Omega\sqrt{\frac{U}{2}}\sum_{k=1}^{M}
\left[ (\alpha_k\partial_{\alpha_k}+\beta_k\partial_{\beta_k})
-({\alpha_k}^*\partial_{{\alpha_k}^*}+{\beta_k}^*\partial_{{\beta_k}^*})
  \right]   \nonumber
\end{equation}
\vspace{-0.5cm}
\begin{equation}\label{Si2_1}
\times i\sqrt{\frac{U}{2}}(n^\prime_k-|n_k|) = -i \Omega  U \sum_{k=1}^{M} n^{\prime\prime}_k
\end{equation}
Considering \eref{Si1_4}, we obtain
\begin{equation}\label{S0}
S_0= \Omega\sum_{j=1}^{M} (2g^2_j-i U n^{\prime\prime}_j)
\end{equation}
Furthermore, $\partial_{\Omega} \underline{\underline{\rm B}}_{i j}=\partial_{\Omega^*}
 \underline{\underline{\rm B}}_{i j}=0$ which, according to \eref{Si}, gives $S^1_i=0$.
Moreover
\begin{equation}\label{Si1_5}
S^2_i= -{\frac{U}{2}}\sum_{j=1}^{2M}  \sum_{k=1}^{2M} (\delta_{kj}\alpha^j\partial_{\alpha^k}\delta_{ij}
-{\delta_{kj}}{\alpha^j}^*\partial_{{\alpha^k}^*}\delta_{ij})\alpha^j \nonumber
\end{equation}
\vspace{-0.6cm}
\begin{equation}\label{Si1_5}
=-{\frac{U}{2}}\alpha^i
\end{equation}
Therefore, we have $S_i= -{\frac{U}{2}}\alpha^i$.
The Stratonovich equations, \eref{StAlpha}, are now
\begin{equation}\label{StraOmega}
d\Omega^{(S)} = \Omega \left(V d\tau + \sum_{k=1}^{2M} g_k dW_k \right) - \frac{S_0}{2} d\tau \nonumber
\end{equation}
\vspace{-0.6cm}
\begin{equation}
= \Omega \left[\left(V -\sum_{j=1}^{M} (g^2_j-i \frac{U}{2} n^{\prime\prime}_j)\right) d\tau + \sum_{k=1}^{2M} g_k dW_k \right]
\end{equation}
\vspace{-0.6cm}
\begin{equation}
d\alpha^{j (S)} = \left[\frac{J}{2} \sum_{i=1}^{2M} {\omega_{ji}} \alpha^i -
 \frac{U}{2}(|n_j|+in^{\prime\prime}_j)\alpha^j + \frac{\mu_e}{2}\alpha^j\right] d\tau \nonumber
\end{equation}
\vspace{-0.6cm}
\begin{equation}
- \frac{S_j}{2} d\tau
+ i \sqrt{\frac{U}{2}}\alpha^j dW_j  \nonumber
\end{equation}
\vspace{-0.6cm}
\begin{equation}
= \left[\frac{J}{2} \sum_{i=1}^{2M} {\omega_{ji}} \alpha^i - \frac{U}{2}(|n_j|
+in^{\prime\prime}_j)\alpha^j + \frac{2\mu_e+U}{4}\alpha^j\right] d\tau \nonumber
\end{equation}
\vspace{-0.6cm}
\begin{equation}\label{StraOmega}
+ i \sqrt{\frac{U}{2}}\alpha^j dW_j
\end{equation}
%\end{widetext}
\begin{acknowledgements}
We would like to thank Peter Drummond, Peter Zoller, Matthias
Troyer, Alexander Mering, Daniel Heinzen, Peter Hannaford, Bryan
Dalton and Chris Vale for helpful discussions. This project is
supported by the ARC Centre of Excellence for Quantum Atom Optics
and a Swinburne University Strategic Initiative fund.
\end{acknowledgements}
\vspace{-0.6cm}
\bibliography{saeed}
\end{document}